\documentclass{aa}  
\usepackage{graphicx}
\usepackage{txfonts}
\usepackage{subfigure}
\usepackage{amsmath}
\usepackage{amssymb}
\usepackage{placeins}

\newcommand{\kms}{\ensuremath{\rm km\,s^{-1}}\xspace}
\newcommand{\Msun}{\ensuremath{\rm M_{\sun}}\xspace}

\newcommand{\Jyb}{\ensuremath{\rm Jy\,beam^{-1}}\xspace}
\newcommand{\vlsr}{\ensuremath{\rm v_{\mathrm{LSR}}}\xspace}

\begin{document}

   \title{Dense clumps survive in the vicinity of R136 in 30 Doradus}

   \author{M. T. Valdivia-Mena\inst{1,2,3}, M. Rubio\inst{1}, V. M. Kalari\inst{4}, H. Salda\~no\inst{5, 6}, A. Bolatto\inst{7}, R. Indebetouw\inst{8, 9}, H. Zinnecker\inst{10}, and C. Herrera\inst{11, 12}}

   \institute{Departamento de Astronom\'ia, Universidad de Chile, Santiago, Chile\\ \email{mariateresa.valdiviamena@eso.org}
        \and
        Max-Planck-Institut f\"ur extraterrestrische Physik, Giessenbachstrasse 1, D-85748 Garching, Germany
        \and
        European Southern Observatory, Karl-Schwarzschild-Strasse 2 85748 Garching bei Munchen, Munchen, Germany
        \and
        Gemini Observatory/NSF’s NOIRLab, Casilla 603, La Serena, Chile
        \and
        Instituto de Investigaciones en Energía no Convencional, Universidad Nacional de Salta, C.P. 4400, Salta, Argentina 
        \and
        Consejo Nacional de Investigaciones Cient\'ificas y T\'ecnicas (CONICET), Godoy Cruz 2290, CABA, CPC 1425FQB, Argentina
        \and
        Department of Astronomy, University of Maryland, College Park, Maryland 20742
        \and
        University of Virginia Astronomy Department, P.O. Box 400325, Charlottesville, VA 22904, USA
        \and
        National Radio Astronomy Observatory, 520 Edgemont Rd., Charlottesville, VA 22903, USA
        \and
        Nucleo de Astroqu\'imica y Astrof\'isica, Universidad Aut\'onoma de Chile, Pedro de Valdivia 425, Providencia, Santiago de Chile, Chile
        \and
        Institut Laue-Langevin, 71 Avenue des Martyrs, 38042 Grenoble, France
        \and
        Institut de Radioastronomie Millimétrique (IRAM), 300 Rue de la Piscine, 38406 Saint-Martin-d’Hères, France}

   \date{}
   
   \titlerunning{Dense clumps survive in the vicinity of R136 in 30 Doradus}
   \authorrunning{M. T. Valdivia-Mena et al.}

 
  \abstract
   {The young massive cluster R136 at the center of 30 Doradus (30 Dor) in the Large Magellanic Cloud (LMC) generates a cavity in the surrounding molecular cloud. However, there is molecular gas between 2 and 10 pc in projection from R136's center. The region, known as the Stapler nebula, hosts the closest known molecular gas clouds to R136.}
   {We investigated the properties of molecular gas in the Stapler nebula to better understand why these clouds survive so close in projection to R136.}
   {We used Atacama Large Millimeter/Sub-millimeter Array 7m observations in Band 7 (345 GHz) of continuum emission, $^{12}$CO and $^{13}$CO, together with dense gas tracers CS, HCO$^+$, and HCN. Our observations resolve the molecular clouds in the nebula into individual, parsec-sized clumps. We determined the physical properties of the clumps using both dust and molecular emission, and compared the emission properties observed close to R136 to other clouds in the LMC.}
   {The densest clumps in our sample, where we observe CS, HCO$^+$, and HCN, are concentrated in a northwest-southeast diagonal seen as a dark dust lane in HST images. Resolved clumps have masses between $\sim 200-2500$ \Msun, and the values obtained using the virial theorem are larger than the masses obtained through $^{12}$CO and $^{12}$CO luminosity. The velocity dispersion of the clumps is due both to self-gravity and the external pressure of the gas. 
   Clumps at the center of our map, which have detections of dense gas tracers ($n_{crit}\sim10^6$ cm$^{-3}$ and above), are spatially coincident with young stellar objects.}
   {The clumps' physical and chemical properties are consistent with other clumps in 30 Dor. We suggest that these clumps are the densest regions of a Molecular Cloud carved by the radiation of R136.}


   \maketitle
%

\section{Introduction\label{sec:introd}}

Massive stars play an important role shaping the interstellar medium (ISM) of a galaxy, through their strong UV radiation stellar winds and eventual explosions as supernovae, and thus, they impact subsequent star formation \citep[e.g.,][]{Shetty2008feedbacksim,Krauss2013feedbacksim,Skinner2015feedbacksim}. The impact of feedback in the properties of Molecular Clouds is twofold: mechanical feedback and radiation pressure can push gas and trigger star formation by concentrating material for further collapse, but the stellar UV radiation destroys surrounding molecular gas and ionizes the medium, disrupting Molecular Clouds and thus slowing down star formation \citep{mckee2007}. Therefore, the study of regions affected by the presence of massive stars is required for a complete picture of star formation in galaxies. 

A unique region to explore the effects of massive stars in star formation is 30 Doradus (30 Dor), also known as the Tarantula Nebula due to its filamentary appearance. 30 Dor is a giant \ion{H}{II} region within the Large Magellanic Cloud (LMC), the nearest dwarf irregular galaxy \citep[at a distance of 50\,kpc,][]{Pietrzynski2013}, characterized by its low metallicity (1/2\,$Z_{\odot}$). 30 Dor is one of the brightest \ion{H}{II} regions in the local universe \citep{kennicutt1984}, consisting of a complex system of filaments and clumps containing hundreds of massive stars \citep[$>15$ \Msun,][]{schneider2018}, with several episodes of star formation in the past 30 Myr \citep[e.g.,][]{Grebel2000AHodge301,Cignoni2016Hodge301,Schneider2018VLTFLAMES, Fahrion202430dorJWST} and where stars are still forming today \citep[e.g.,][]{gruendl2009, Walborn2013spitzer, Kalari2014VLT-Flames, Ksoll2018PMSstars, Nayak2023YSOtarantula}.

30 Dor hosts the young massive cluster (YMC) R136, a $\sim1-2$ Myr old compact cluster \citep{Crowther2016stellarpropsR136}, with an extremely high density \citep[about $10^{4}$ \Msun pc$^{-3}$,][]{Selman2013densityR136} and containing several stars more massive than 150 \Msun \citep{Crowther2010massivestars, Brands2022mostmassivestars}. 
This YMC contributes to about a quarter of the ionizing flux and about a fifth of the total mechanical feedback in the whole nebula \citep{Doran2013VLT-Flames, Bestenlehner2020R136ebudget}.
According to most theoretical predictions \citep[][and references therein]{dale2012}, the photoionising luminosity (of 10$^{51}$ ph\,s$^{-1}$) is sufficient to evaporate any dense molecular gas surrounding the cluster ($\leq$10-15\,pc). 
Feedback from the R136 stars have indeed generated a cavity with an apparent radius of $\sim10$ pc by sweeping molecular gas, forming elongated, pillar-like structures, through its ionizing radiation \citep{chu1994, johansson1998, Pellegrini2010}.

Surprisingly, there is cold molecular emission near R136, between 2 and 10 pc in projection \citep{Rubio2009, kalari2018}. 
\cite{Rubio2009} found dense molecular gas emission (10$^6$\,cm$^{-3}$) associated with a young massive star, IRSW-127 \citep{rubio1998}. \cite{kalari2018} investigated further this region through $^{12}$CO $J = 2-1$ emission and found three molecular clouds, ``knots'', with a total mass of 2$\times10^{4}\,M_{\odot}$. The molecular gas shows several velocity components ranging from  $\sim$ 235 to 250\,km\,s$^{-1}$ with complex velocity profiles showing many components and suggesting these molecular clouds further divide into smaller clumps. The cold molecular gas distribution spatially coincides with a dark lane seen at optical wavelengths and with parsec scale knotty structures seen in excited H$_2$ gas in the near-infrared \citep{kalari2018}, suggesting that these knots are being photoionized. The velocity of the cold molecular gas and the spatial structure of the excited warm molecular gas both suggest that the cold gas is located close enough to R\,136 that the radiation output from the R\,136 cluster should be photoionizing the molecular clouds. 

In this work, we present ALMA observations of the molecular gas structure at 4.7\arcsec (1.1 pc at the adopted distance to the LMC) resolution towards the vicinity of R136 in 30 Dor. We resolve the ``knots'' seen by \cite{kalari2018} into sub-parsec size clumps and using different molecules detected in the 345 GHz window, we determine the physical conditions of the gas exposed to the ionizing radiation of R136. The paper is organized as follows. 
In Sect. \ref{sec:obs}, we describe the ALMA observations used and how we processed the images obtained. Section \ref{sec:clumpidmethods} describes the methods used to identify clumps in molecular and continuum emission. In Section \ref{sec:results} we report the physical properties of the clumps based on the $^{12}$CO $J=3-2$ clumps detections (Sect. \ref{sec:physprops}), $^{13}$CO $J=3-2$ emission (Sect. \ref{sec:columndensities}), CS $J=7-6$, HCO$^+$ $J=4-3$, and HCN $J=4-3$ emission (Sect. \ref{sec:mollinesint}) and dust emission (Sect. \ref{sec:propsdust}).
In Sect. \ref{sec:coandcontcomparison} we compare the clump properties between gas and dust emission. We discuss our results in Sect. \ref{sec:discussion}. We summarize our findings in Sect. \ref{sec:summary}.

\section{Observations and data reduction\label{sec:obs}}

\begin{table*}[ht]
\caption{\label{tab:cubeprops}Characteristics of the resulting line emission cubes.}
\centering
\begin{tabular}{@{}ccccccccc@{}}
\hline\hline
Molecule     & Transition  & Rest frequency  & $n_{\mathrm{crit}}$ & Beam FWHM  & PA  & $\Delta v_{chan}$ & Pixel size  & $\sigma$  \\

& & (GHz) & (cm$^{-3}$) & (\arcsec) & (deg) & (km s$^{-1}$) & (\arcsec) &  (mK)  \\ \hline
$^{12}$CO & $J=3-2$ & 345.796  &  $3.9\times10^3$ & 4.6 x 4.0  & 54 & 0.211  & 0.7  & 47  \\
$^{13}$CO & $J=3-2$ & 330.587 & $3.4\times10^3$ & 4.7 x 4.0   & 79  & 0.221  & 0.7    & 51    \\
CS & $J=7-6$    & 342.883  & $^*4.4\times10^{6}$     & 4.6 x 4.0   & 57  & 0.213  & 0.7   & 34   \\
HCO$^+$ & $J=4-3$   & 356.734  & $^*3.2\times10^{6}$  & 4.7 x 3.6 & 77   & 0.410   & 0.7   & 30   \\
HCN & $J=4-3$  & 354.505  & $^*2.3\times10^{7}$  & 4.7 x 3.6 & 76 & 0.413  & 0.7  & 26   \\ \hline
\end{tabular}
\tablefoot{PA is the angle between the x axis and the major axis of the beam, counterclockwise. $n_{\mathrm{crit}}$ is obtained assuming $T=20$ K. \tablefoottext{*}{\cite{Shirley2015}}.}
\end{table*}

\begin{figure}
    \centering
    \includegraphics[width=1\linewidth]{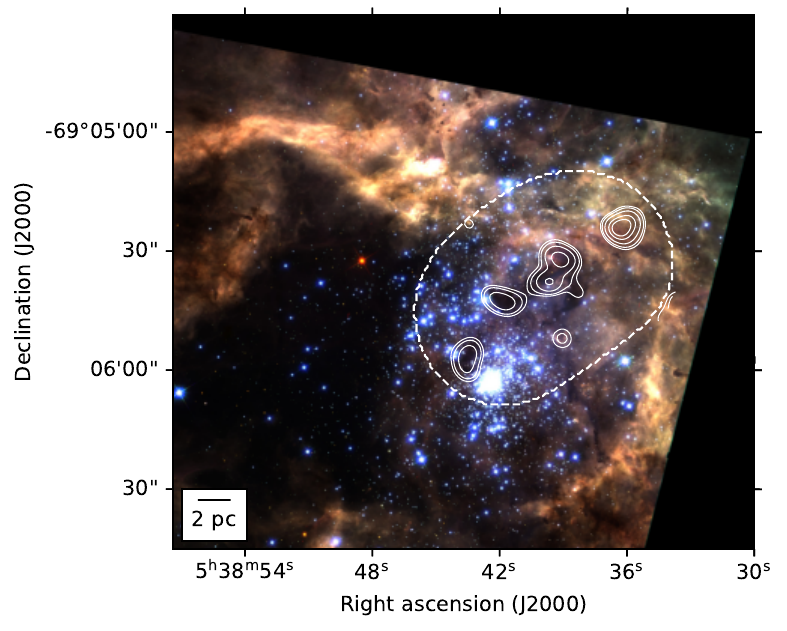}
    \caption{Moment 0 contours of the $^{12}$CO $J=3-2$ ALMA observations used in this work, plotted over a B/V/I H$\alpha$ image from the Hubble Space Telescope (HST). Contours correspond to 3, 5, 10, 20 and 30 times the rms ($=4.92$ \Jyb \kms) of the $^{12}$CO integrated image between 217 and 285 \kms. The dashed line marks the area covered by the ALMA observations.}
    \label{fig:hst-footprint}
\end{figure}

Observations were taken with the Atacama Large Millimeter/Sub-millimeter Array (ALMA) during Cycle 5, using the 7m ALMA Compact Array (ACA, also known as the Morita array) under project 2017.1.00368.S (PI M. Rubio). The observations consist of two Band 7 correlator setups. The first setup was tuned to observe the $^{12}$CO $J=3-2$, $^{13}$CO $J=3-2$, and CS $J=7-6$ molecular lines, each in one spectral window (spw), with an additional spw to observe continuum emission with a total bandwidth of 2 GHz. The second setup was tuned to observe the HCO$^+$ $J=4-3$ and HCN $J=4-3$ transitions, each in an individual spw, and two additional spw for continuum. The phase center of all data is $\alpha=05^h38^m39.95^s$, $\delta= -69^{\circ}05^m40.33^s$ (J2000) and both correlator setups cover a field of view of approximately 1.2'$\times$1.2'. The spatial resolution is approximately 4.7\arcsec for all cubes, which corresponds to 1.1 pc at a distance of 50 kpc. The maximum recoverable scale (MRS) for the first setup is 19.1\arcsec (4.6 pc) and for the second setup is 18.6\arcsec (4.5 pc). As we only have one transition for each molecule, we refer to $^{12}$CO $J=3-2$, $^{13}$CO $J=3-2$, CS $J=7-6$, HCO$^+$ $J=4-3$, and HCN $J=4-3$ emission as $^{12}$CO, $^{13}$CO, CS,  HCO$^+$, and  HCN, unless otherwise stated. The footprint of the first setup observations is plotted with respect to an HST image of 30 Dor, highlighting the location of the R136 cluster, in Fig. \ref{fig:hst-footprint}.

\begin{figure*}
	\centering
	\includegraphics[width=0.95\textwidth]{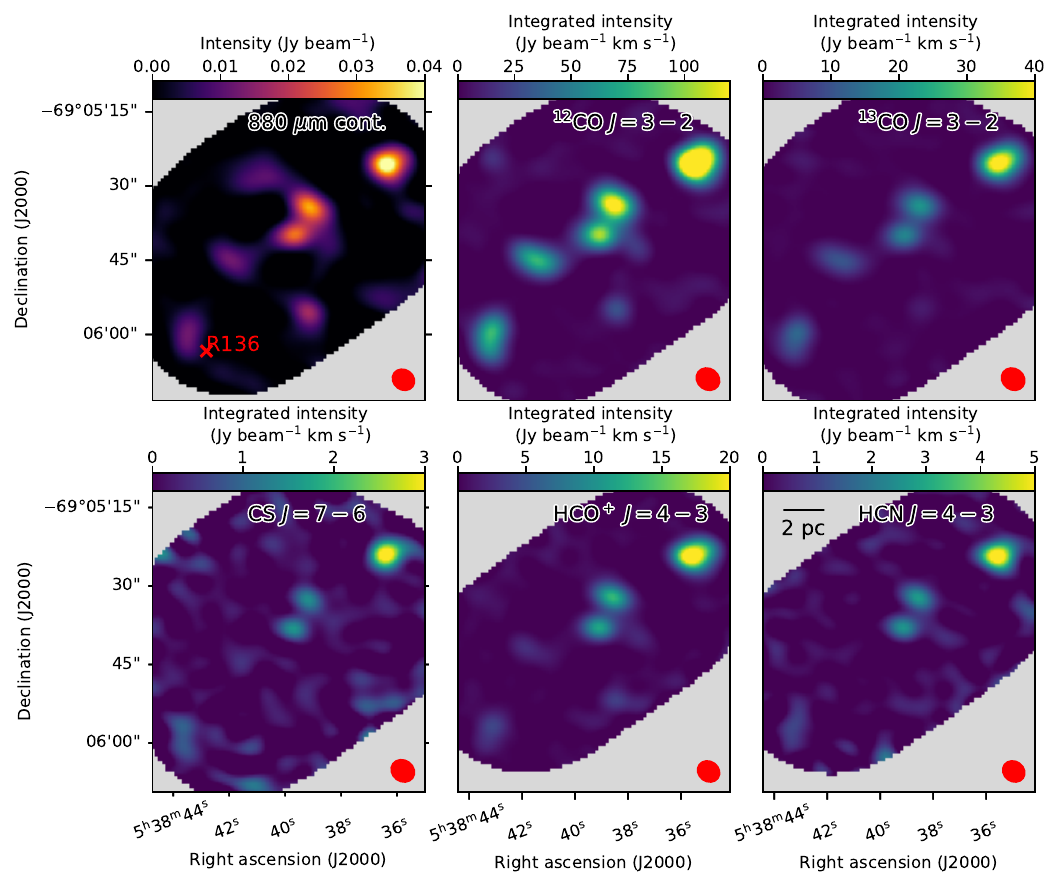}
	\caption{ALMA 0.88 mm continuum image of 30 Dor in the vicinity of R136, together with the moment 0 maps of the molecules observed in this work. The moment 0 images are made integrating each molecular emission between 217 and 285 \kms. Each molecule and image is labeled in the top part. The black scale-bar represents a 2 pc length. The red ellipse represents the beam size. In the top left panel (continuum), the red cross marks the center of the R136 YMC. } 
	\label{fig:all-mom0}
\end{figure*}

We obtained the calibrated data using the Common Astronomy Software Applications package (CASA) Pipeline v.5.4.0.70, using the standard scripts provided by ALMA Operations Support Facility (OSF) with the delivered raw data. Imaging was performed in CASA v.5.6.1. For the molecular line cubes, we first ran the \texttt{uvcontsub} task to subtract continuum emission in the molecular line maps via a 0th order fit to line-free channels. Then we used the \texttt{tclean} task to deconvolve and clean the data cubes. We deconvolved with Briggs weighting using a robust parameter of 0.5. We used the Hogbom CLEAN algorithm with manually-applied masking to reduce negative bowls of emission in line-free regions.  The final properties of the data cubes (beam size, position angle, spectral resolution $\Delta v_{chan}$, pixel size, rms $\sigma$ and rest frequency $\nu_{rf}$) are in Table \ref{tab:cubeprops}. Integrated emission images of all molecules are shown in Fig. \ref{fig:all-mom0}.

We produced one continuum image by combining the continuum spw from both setups. We first concatenated the spws together in one calibrated file using the \texttt{concat} routine with a frequency tolerance of 10 MHz. We flagged the channels which contain line emission. We manually flagged channels that present increased amplitude due to the atmospheric transmission to improve the noise level of the final continuum image. We deconvolved the concatenated data using multi-frequency synthesis, implemented in the \texttt{tclean} task, with natural weight. We do an interactive Hogbom CLEAN to apply a manual mask to the dirty image. The final continuum image is shown in the top left panel of Fig.~\ref{fig:all-mom0}. It has a central frequency of 338.5 GHz (0.88 mm), a total bandwidth of 13.5 GHz, a resolution of 4.7\arcsec $\times$ 3.9\arcsec (position angle of 69$^{\circ}$), a pixel size of 0.7\arcsec, and an RMS of 4 m\Jyb.

We applied primary beam correction to all data products using the primary beam response given by the \texttt{tclean} task after the CLEAN process. In the case of the line cubes, each channel was divided by the primary beam response. Because of this correction, $\sigma$ is not uniform: as the primary beam response is lower towards the edges than the center, the noise increases radially.

\section{Clump identification methods\label{sec:clumpidmethods}}

Figure \ref{fig:all-mom0} shows that emission consists of clumpy structures in both molecular line and continuum emission, mostly concentrated in a diagonal that goes from the northwest to southeast. A visual inspection of the line cubes reveals that all molecules have emission between 210 and 290 \kms. $^{12}$CO and $^{13}$CO maps also contain smaller, less bright clumps toward the south of this diagonal. We identified the individual emission structures in the line cubes using the cloud identification algorithm CPROPS \citep{rosolowsky2006, Rosolowsky-Leroy2011CPROPScode}, together with Gaussian fitting of the $^{12}$CO map. Individual emission structures were identified and characterized in the continuum image through visual inspection and aperture photometry. 
Following the terminology used in \cite{Wong2022-30Dorclumps}, we refer to the individual structures as clumps as we find structures comparable to their clumps in size ($\sim1$ pc).

\subsection{$^{12}$CO clump identification with CPROPS\label{sec:cprops}}

\begin{figure}
    \centering
    \includegraphics[width=0.485\textwidth]{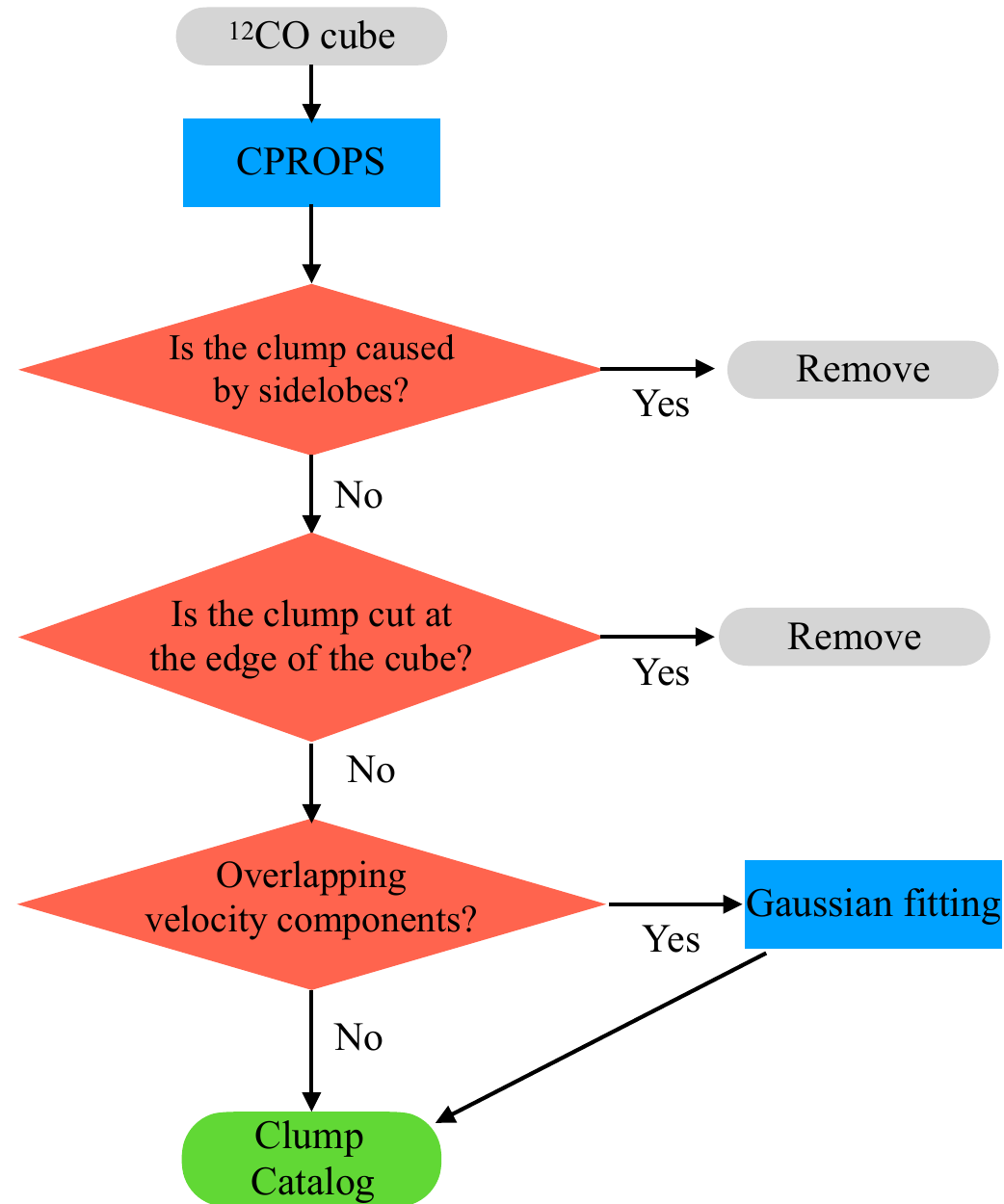}
    \caption{Flowchart of the process to identify the clumps in the ALMA $^{12}$CO data.}
    \label{fig:flowchart}
\end{figure}

We first identified and characterized $^{12}$CO line emission, as it presents the largest line intensities of our sample. Figure \ref{fig:flowchart} shows the steps taken to obtain the clump catalog from the $^{12}$CO emission cube. We then ran CPROPS with the other molecular line cubes investigated in this work (Table \ref{tab:cubeprops}), using the parameters that worked best on the $^{12}$CO line map, and cross-matched the results of all molecules to the $^{12}$CO results. The resolution of our data (Table \ref{tab:cubeprops}) allowed us to resolve clumps with diameters down to approximately 1 pc.

We describe the parameters we used in CPROPS in the following. We used the IDL implementation of CPROPS\footnote{https://github.com/low-sky/cprops} \citep{Rosolowsky-Leroy2011CPROPScode}. Detailed descriptions of the algorithm are in \cite{rosolowsky2006, Rosolowsky-Leroy2011CPROPScode}. CPROPS identifies emission as three-dimensional ``islands'' over a certain noise level and decomposes them into single clumps, assigning each pixel in the line cube to a clump or as background noise. We set the parameters THRESH $=3\sigma$ and EDGE $=1.5\sigma$ ($\sigma$ from Table \ref{tab:cubeprops}) to define the initial mask, from which CPROPS will decompose the emission. We used the /NONUNIFORM flag to account for the non-uniform noise in our line cube. We set the minimum area for a clump to be included with the MINAREA parameter, which we set to $0.5$ resolution elements (beam area). We set the MINPEAK parameter, the minimum peak value of a potential clump, to $3\sigma$. We set the minimum number of contiguous channels with the MINVCHAN parameter, which we set to MINVCHAN $=3$ channels (which equals 0.6 km s$^{-1}$). We use the ECLUMP variation of the CPROPS algorithm, which allows emission shared within a single brightness contour level by two clumps to be assigned into the clump with the closest peak, using the CLUMPFIND algorithm \citep{williams1994}. A detailed explanation of what the ECLUMP variation does can be found in the CPROPS user guide. Finally, we set BOOTSTRAP $=1000$ so CPROPS does 1000 bootstrap iterations to estimate the uncertainties in the detection properties.

We manually inspected the results for the $^{12}$CO detections by CPROPS to check for false clumps caused by artifacts in the interferometric data (sidelobes). The program identified a total of 49 clumps, 35 of which have a signal-to-noise ratio ($S/N$) higher than 5 (around 70\% of the total). We first check if the identified clumps are emission sidelobes caused by brighter clumps in the vicinity. All clumps with $S/N< 5$ are false detections caused by the superposition of emission arising from the sidelobes of stronger neighboring clumps. From the 35 clumps with $S/N> 5$, 4 of them are false detections caused by sidelobes. We also discard 8 clumps which are at the borders of the map because we cannot characterize their structure. For the remaining 23 clumps, we determined that there are 19 clumps identified by CPROPS which are consistent with our visual inspection of the $^{12}$CO cube. The remaining four clumps are actually five individual clumps that are blended. We kept the 19 clumps that are consistent with what we observe and use Gaussian fitting to obtain the properties of the remaining clumps. We describe the Gaussian fitting process in Appendix \ref{sec:cpropssecondstep}.

\begin{table}[!ht]
\caption{\label{tab:cpropsresults}$^{12}$CO Clump detections via CPROPS }
	\centering
	\begin{tabular}{cccccc}
		\hline\hline
		ID &    $\alpha$ &     $\delta$ & $T_{peak}$  & $v$  &     $\Delta v$  \\
		& (J2000)  & (J2000) & (K) & (km s$^{-1}$) & (km s$^{-1}$) \\
		\hline
		1 & 05:38:36.1 & -69:05:46.9 & 1.31 & $271.2$ & $1.3\pm0.2$ \\
		2 & 05:38:36.2 & -69:05:24.1 & 25.74 & $250.2$ & $4.2\pm0.2$ \\
		3 & 05:38:36.5 & -69:05:31.5 & 0.60 & $219.2$ & $1.9\pm0.2$ \\
		4 & 05:38:36.7 & -69:05:29.9 & 0.25 & $221.1$ & $1.1\pm0.5$ \\
		5 & 05:38:36.9 & -69:05:50.7 & 0.50 & $279.8$ & $2.5\pm0.4$ \\
		6 & 05:38:37.5 & -69:05:48.7 & 0.28 & $255.0$ & $2.6\pm0.5$ \\
		7$^{a}$ & 05:38:37.6 & -69:05:33.3 & 0.28 & $224.7$ & $2.6\pm0.2$ \\
		8$^{a}$ & 05:38:37.6 & -69:05:33.6 & 0.76 & $227.6$ & $2.1\pm0.1$ \\
		9 & 05:38:38.0 & -69:05:44.6 & 0.31 & $228.5$ & $2.0\pm0.3$ \\
		10 & 05:38:38.3 & -69:05:39.5 & 1.75 & $246.8$ & $2.5\pm0.3$ \\
		11 & 05:38:38.5 & -69:05:40.5 & 1.30 & $232.5$ & $3.6\pm0.2$ \\
		12 & 05:38:38.7 & -69:05:37.5 & 2.08 & $237.4$ & $2.9\pm0.2$ \\
		13 & 05:38:38.9 & -69:06:01.7 & 1.00 & $276.2$ & $2.7\pm0.3$ \\
		14 & 05:38:38.9 & -69:05:45.4 & 1.52 & $241.3$ & $2.1\pm0.3$ \\
		15 & 05:38:39.1 & -69:05:52.0 & 4.23 & $247.5$ & $3.6\pm0.3$ \\
		16$^{a}$ & 05:38:39.2 & -69:05:31.8 & 25.13 & $244.4$ & $3.1\pm0.0$ \\
		17$^{a}$ & 05:38:39.6 & -69:05:37.7 & 9.20 & $245.4$ & $4.3\pm0.3$ \\
		18$^{a}$ & 05:38:39.8 & -69:05:37.5 & 8.23 & $241.3$ & $2.4\pm0.1$ \\
		19 & 05:38:40.0 & -69:06:04.6 & 0.64 & $267.5$ & $2.2\pm0.4$ \\
		20 & 05:38:40.4 & -69:06:06.7 & 0.76 & $273.2$ & $4.6\pm0.5$ \\
		21 & 05:38:41.7 & -69:05:24.8 & 0.53 & $254.2$ & $1.8\pm0.3$ \\
		22 & 05:38:41.8 & -69:05:42.3 & 13.15 & $237.7$ & $3.0\pm0.2$ \\
		23 & 05:38:43.5 & -69:05:57.0 & 11.49 & $235.9$ & $3.5\pm0.2$ \\
		24 & 05:38:44.1 & -69:05:59.1 & 0.49 & $227.8$ & $2.9\pm0.4$ \\
		\hline
	\end{tabular}
	\tablefoot{The velocity FWHM is corrected for sensibility and resolution bias.
    \tablefoottext{a}{Characterized using a manual method, described in Appendix \ref{sec:cpropssecondstep}.}}
\end{table}

\begin{figure*}
	\centering
	\includegraphics[width=0.95\textwidth]{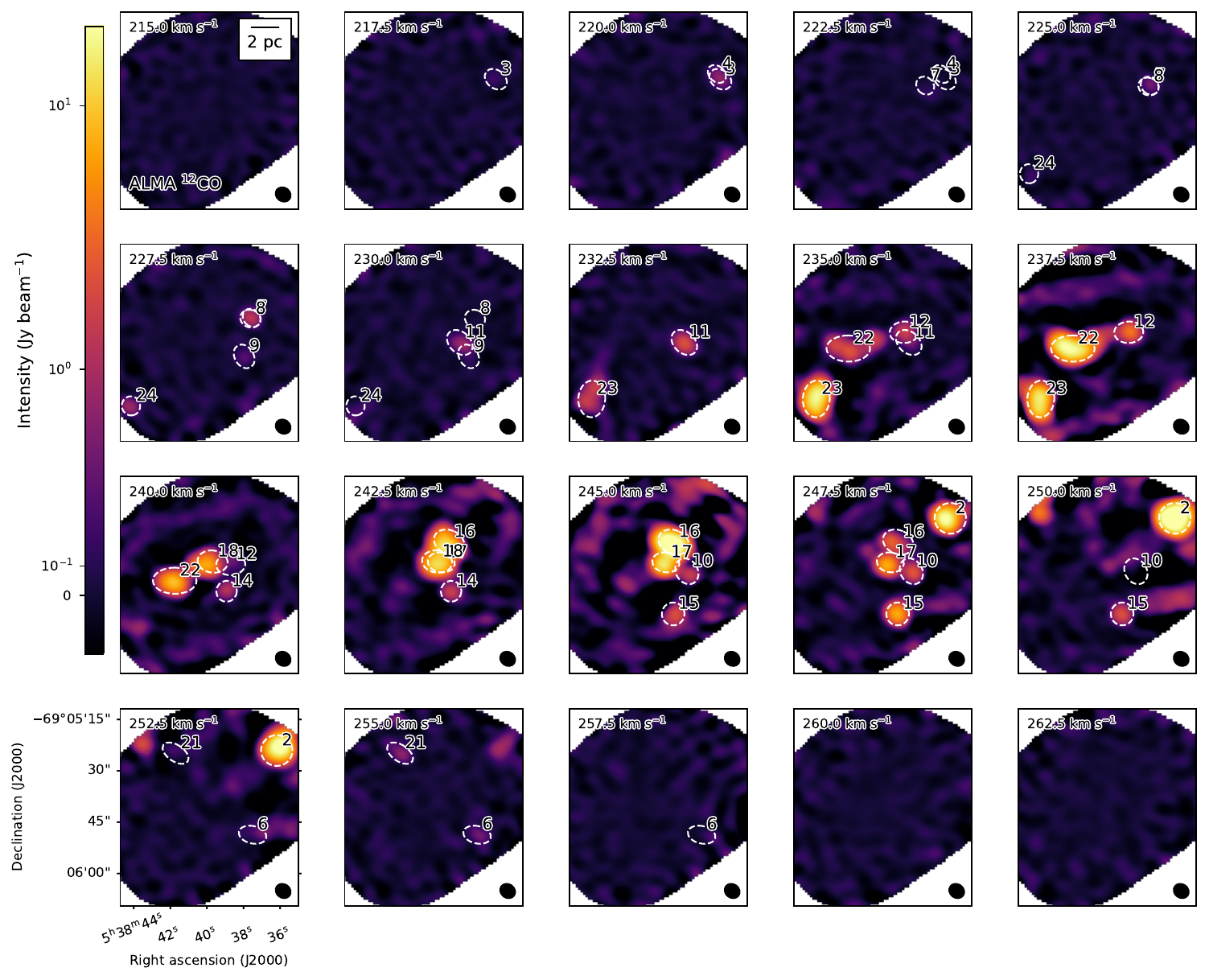}
	\caption{$^{12}$CO channel maps between 215 and 265 \kms with the clumps found using CPROPS. The major and minor radii of the ellipses correspond to 1.91$\sigma_{maj}$ and 1.91$\sigma_{min}$, where $\sigma_{maj}$ and $\sigma_{min}$ are the extrapolated (not deconvolved) second moments of emission along the major and minor axis of the clumps, obtained using CPROPS except for clumps 7, 8, 16, 17 and 18, which are obtained as described in Appendix \ref{sec:cpropssecondstep}. The black ellipse in the lower right corner represents the beam size. The scale-bar represents a length of 2 pc. \label{fig:chanmap-cprops} }
\end{figure*}

The final catalog of  $^{12}$CO clumps consists of 24 sources, listed in Table \ref{tab:cpropsresults}. We plot the detections in Figure \ref{fig:chanmap-cprops} with ellipses that represent the extrapolated (but not deconvolved) radii along the major and minor axes of the clump (see Sect. \ref{sec:physprops}), as delivered by CPROPS. Each ellipse is plotted in three channel maps, centered in the channel map that has the closest velocity to the clump's $v$. The clumps which have been characterized by the Gaussian fitting method correspond to clumps no. 7, 8, 16, 17 and 18. 

\subsection{Clump identification in all other molecular emission\label{sec:molemissionCPROPS}}

\begin{figure}[!ht]
	\centering
	\includegraphics[width=0.9\hsize]{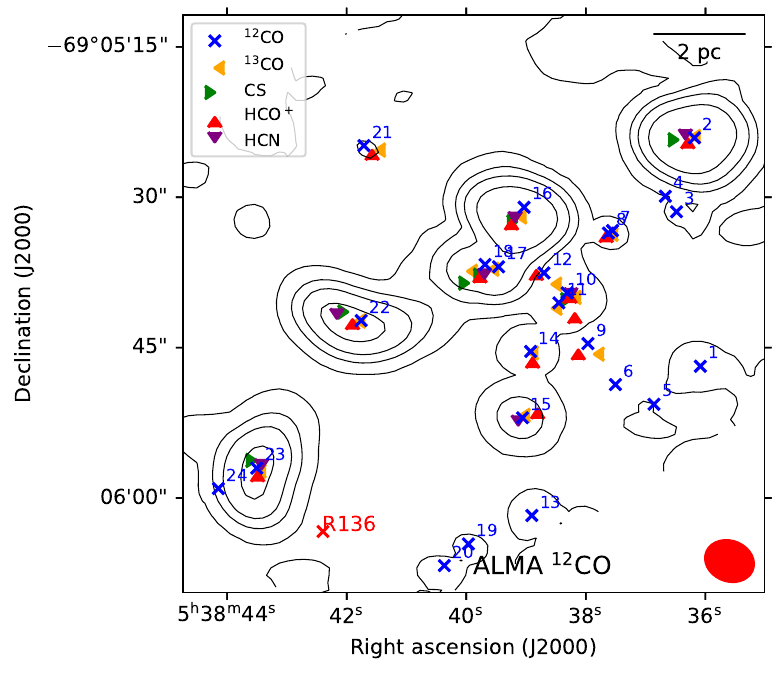}
	\caption{\label{fig:moleculepos}Central position of the clumps found in $^{12}$CO, $^{13}$CO, CS, HCO$^{+}$ and HCN, plotted over contours of the peak main beam temperature of $^{12}$CO. Contours correspond to 5, 10, 100, and 200 times the rms of the $^{12}$CO line emission (47 mK). Blue crosses mark the central position of the clumps found in the $^{12}$CO in Sect. \ref{sec:cprops}. Yellow arrowheads point towards the central position of $^{13}$CO clumps. Green arrowheads point towards the central position of CS clumps. Red arrowheads point towards the central position of HCO$^{+}$ clumps. Purple arrowheads point towards the central position of HCN clumps. The red ellipse in the lower right corner represents the beam size.}
\end{figure}

We identified the molecular clump emission in $^{13}$CO, CS, HCO$^{+}$, and HCN using CPROPS with the same parameters as those used to identify the $^{12}$CO clumps (Sect. \ref{sec:cprops}). We followed almost the same procedure as the one shown in Fig. \ref{fig:flowchart}, except we did not separate overlapping velocity components. We manually inspected the results to leave out false detections produced by artifacts and include detections in the positions of $^{12}$CO clumps which are not detected by CPROPS. We list the CPROPS detections and their properties (position, central velocity, FWHM and sizes) in Appendix \ref{appendix:cpropsmolecules}. In summary, we detected 13 clumps in $^{13}$CO, located between 225 and 255 \kms, 8 in CS between 233 and 253 \kms, 12 in HCO$^{+}$ between 225 and 255 \kms, and 6 in HCN between 233 and 253 \kms. These clumps are plotted together with $^{12}$CO clumps in Figure \ref{fig:moleculepos}. Figure \ref{fig:moleculepos} plots the position of each clump in the different molecules detected. In general, these molecules are found towards the brightest $^{12}$CO but not all clumps have detectable emission in all the lines, as shown in Fig. \ref{fig:moleculepos}. We used the $^{12}$CO clump ID to identify the clumps in the different molecules in Appendix \ref{appendix:cpropsmolecules}.

We found emission of all the molecular species associated only to six $^{12}$CO clumps (no. 2, 10, 16, 17, 22, and 23, see Figure \ref{fig:moleculepos}). These are the brightest $^{12}$CO, all located along the diagonal which coincides with dust emission in the optical. The highest $T_{peak}$ values for all molecular species are found in clump no.2 (according to the tables in Appendix \ref{appendix:cpropsmolecules}), which is the farthest away in projection from R136. 
CS and HCN are detected in the strongest CO clumps located in the northwest-southeast diagonal. These detections imply these are very dense clumps, as CS $J=7-6$ and HCN $J=4-3$ have high critical densities, $n_{crit}\sim10^7$ cm$^{-3}$ and $n_{crit}\sim10^8$ cm$^{-3}$, respectively. To determine the volume density and temperature of the clumps, non-LTE modeling using additional line transitions is required, which is out of the scope of this work. %
In particular, $^{13}$CO emission is detected in 14 $^{12}$CO clumps, 
mainly in the northwest-southeast diagonal cloud structure but also in clumps no. 14, 15 and 21, outside  of this diagonal. The sizes of the $^{13}$CO clumps are similar to the sizes of the $^{12}$CO (Appendix \ref{appendix:cpropsmolecules}), except for clump no. 2 which has a smaller minor axis and therefore, a smaller radii in comparison with its $^{12}$CO $J=3-2$ counterpart ($R_{dc}=0.42\pm0.06$ in $^{13}$CO versus $R_{dc}=0.71\pm0.10$ in $^{12}$CO). 

\subsection{Clump identification in continuum emission \label{sec:contradii}}

Continuum emission shows a similar distribution as the $^{12}$CO integrated molecular line emission (Fig. \ref{fig:all-mom0}). There are continuum emission sources concentrated along the diagonal in northwest-southeast direction and at least one continuum source at the south part of the image. Emission is brightest in the northwest corner of the image and becomes less bright towards the southeast, following the trends observed in all molecular emission.

\begin{figure}[ht]
	\centering
	\includegraphics[width=0.9\hsize]{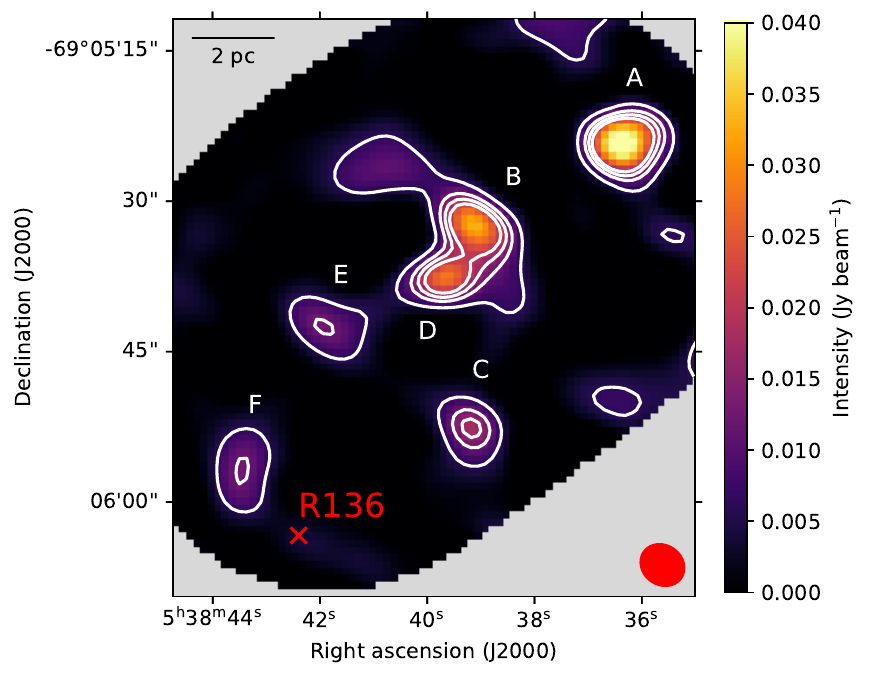}
	\caption{\label{fig:continuumALMA}ALMA 0.88 mm continuum image of the vicinity of R136. White contours represent 1.5, 3, 4, and 5 $\sigma$ levels ($\sigma=4$ mJy/beam), with labels indicating the contours that correspond to clumps A, B, C, D, E, and F. The red ellipse in the lower right corner represents the beam size. The scale-bar in the upper left corner represents a 2 pc length.}
\end{figure}

We detected individual sources in continuum emission and then obtained their sizes and fluxes. For this, we first identified the sources in the ALMA 0.88 mm continuum image through visual inspection. We considered emission over 3$\sigma$ in the continuum image as a detection, where $\sigma = 4$ m\Jyb is the rms of the 0.88 mm continuum image. Using this criterion, we identified 6 sources in the image, which we name A to F in increasing right ascension. The continuum sources are shown labeled in the ALMA continuum image in Fig. \ref{fig:continuumALMA}. Two of these sources, B and D, are close enough to each other so that their emissions share the same 3$\sigma$ contour. All of these sources coincide with one or more $^{12}$CO clumps from Sect. \ref{sec:molemissionCPROPS}: source A coincides with clump no. 2, source B with clump no. 16, source C with clump no.15, source D with clumps no. 17 and no.18, source E with clump no. 22 and source F with clump no. 23. We refer to these continuum sources as clumps as well from now on.

We obtained the areas, radii and flux of each of the detected clumps. 
We defined the area of a clump in the continuum image as the area inside the 1.5$\sigma$ contour of each detected source. These contours, together with the 3, 4, and 5$\sigma$ contours are shown in Figure \ref{fig:continuumALMA}. There are other 1.5$\sigma$ contours that do not contain emission over 3$\sigma$ within them. We do not consider these emissions as clumps as their $S/N$ is lower than 3, and thus these could be produced by sidelobes from the interferometric image. Then, we calculated the area of the clump $A$ as:
\begin{equation}
	A = N_{pix}  d_{pix}^2, \label{eq:areacont}
\end{equation}
where $N_{pix}$ is the number of pixels inside the 1.5$\sigma$ contour of the clump and $d_{pix}$ is the diameter of a pixel in pc. Emission from clumps B and D share the same contours up to a 5$\sigma$ level, so to determine their area, we modeled the emission from both clumps as two elliptical Gaussians. We used the \texttt{astropy.modelling} package to find the best fit for this region of the image. Afterwards, we generated an image with only one of the Gaussians and found the area of the isolated clump.

We calculated the equivalent circular radii $R_{eq}$ of the clumps, using $A$ (Eq. \ref{eq:areacont}): 
\begin{equation}
	R_{eq} = \sqrt{\frac{A}{\pi}}. \label{eq:eqradii}
\end{equation}
We also calculated the deconvolved equivalent radii:
\begin{equation}
	R_{eq, dc} = \sqrt{R_{eq}^2 - R_{beam}^2},\label{eq:deconvolvedeqradii}
\end{equation}
where $R_{beam}$ is the geometric mean of the radii along the major and minor axes of the beam. This formula is equivalent to Equation 9 from \cite{rosolowsky2006} when applied to a circular clump. For the ALMA 0.88 mm continuum image, the beam radii is $R_{beam}=0.84$ pc. If the equivalent radii of a clump is smaller than $R_{beam}$, we adopted $R_{eq, dc}=R_{beam}$ as an upper limit to the deconvolved radius. The resulting areas and equivalent radii (deconvolved and not deconvolved) for the clumps in the continuum image are in Table \ref{tab:radiicont}. Clumps C, E and F are unresolved.

\begin{table*}
\caption{\label{tab:radiicont} Properties of the clumps derived from the ALMA 0.88 mm image.}
	\centering
	\begin{tabular}{cccccccc}
		\hline\hline
		ID & $\alpha$ &     $\delta$ & Area & $R_{eq}$ & $R_{eq,dc}$ & $r_{ap}$ & $S_{880}$  \\
		& (J2000) & (J2000) & (pc$^2$) & (pc) & (pc) & (\arcsec) &  (mJy) \\
		\hline
		A & 05:38:36.3 & -69:05:24.3 & 3.60 & $1.07\pm0.13$ & $0.66\pm0.11$ & 4.90 & $61.17\pm7.66$ \\
		B & 05:38:39.1 & -69:05:32.6 & 3.69 & $1.08\pm0.19$ & $0.69\pm0.15$ & 5.00 & $49.10\pm9.39$ \\
		C & 05:38:39.2 & -69:05:52.7 & 1.96 & $0.79\pm0.25$ & $<0.84$  & 3.90 & $19.35\pm6.10$ \\
		D & 05:38:39.7 & -69:05:37.8 & 2.76 & $0.94\pm0.19$ & $0.42\pm0.14$ & 6.00 & $32.86\pm8.29$ \\
		E & 05:38:41.9 & -69:05:42.5 & 1.96 & $0.79\pm0.35$ & $<0.84$  & 4.20 & $17.61\pm6.57$ \\
		F & 05:38:43.4 & -69:05:57.0 & 2.13 & $0.82\pm0.37$ & $<0.84$  & 4.30 & $18.63\pm6.73$ \\
	\end{tabular}
	
\end{table*}

We determined the total flux coming from the clumps found in the ALMA 0.88 mm continuum image using aperture photometry with background sky subtraction. The aperture of each clump is a circle that encloses the area found in Sect. \ref{sec:contradii} (the 1.5$\sigma$ contour). We defined the center of the aperture as the center of each clump. The radii of each aperture $r_{ap}$ and their central positions are listed in Table \ref{tab:radiicont}. The background emission flux was obtained by taking the median intensity at a sample of apertures with a radius of $5\arcsec$, which do not present emission in the ALMA 0.88 mm continuum. Background emission represented $<5\%$ of the flux within the clumps' apertures. This median was then multiplied by the aperture area of each clump to determine the background emission. The flux density of each source is the flux present inside the aperture minus the background emission. The uncertainties in the fluxes are the photometric errors inside the aperture area:
\begin{equation}
	\epsilon = \sigma\sqrt{N_{beams}}
\end{equation}
where $\sigma$ is the rms of the image (4 m\Jyb) and $N_{beams}$ is the number of beams inside the aperture area. To obtain the flux coming from sources B and D, we used the Gaussian models of each source for the aperture photometry. The total fluxes in the continuum image $S_{880}$ in mJy for each clump are listed in Table \ref{tab:radiicont}.

\section{Results and analysis\label{sec:results}}

\subsection{Physical properties of clumps near R136 based on $^{12}$CO\label{sec:physprops}}

\begin{table*}[ht]
\caption{\label{tab:propclouds}Physical properties of the $^{12}$CO clumps in 30 Dor.}
\centering
\begin{tabular}{ccccccc}
\hline\hline
ID & $L_{^{12}CO}$  & $R$ & $M_{vir}$ &  $M^{CO}_{\mathrm{gas}}$ & $\Sigma_{\text{H}_2}$ & $\frac{M_{vir}}{M^{CO}_{\mathrm{gas}}}$ \\
 & (K km s$^{-1}$ pc$^{-2}$) & (pc) & ($M_{\odot}$) & ($M_{\odot}$) & (M$_{\odot}$ pc$^{-2}$ ) & \\  \hline
1 & $2.24\pm0.34$ & $<0.84^{**}$ & $<266.8\pm91.9$ & $9.4\pm3.6$ & $>4.2\pm2.3$ & $<28.4\pm14.7$ \\
2 & $268.56\pm22.11$ & $0.71\pm0.10$ & $2414.2\pm382.2$ & $1128.0\pm413.4$ & $712.0\pm297.9$ & $2.1\pm0.9$ \\
3 & $1.54\pm0.16$ & $0.13\pm0.02^*$ & $88.0\pm21.8$ & $6.5\pm2.4$ & $120.0\pm53.2$ & $13.6\pm6.1$ \\
4 & $0.18\pm0.04$ & $<0.84^{**}$ & $<194.8\pm163.2$ & $0.7\pm0.3$ & $>0.3\pm0.3$ & $<262.8\pm246.5$ \\
5 & $1.34\pm0.13$ & $<0.84^{**}$ & $<1028.6\pm316.7$ & $5.6\pm2.1$ & $>2.5\pm1.3$ & $<182.5\pm87.8$ \\
6 & $0.84\pm0.20$ & $0.36\pm0.11^*$ & $457.0\pm187.9$ & $3.5\pm1.5$ & $8.7\pm5.3$ & $129.3\pm76.8$ \\
7 & $1.03\pm0.09$ & $<0.84^{**}$ & $<1086.5\pm172.6$ & $4.3\pm1.6$ & $>1.9\pm0.8$ & $<251.5\pm101.0$ \\
8 & $1.78\pm0.08$ & $<0.84^{**}$ & $<714.3\pm45.4$ & $7.5\pm2.7$ & $>3.4\pm1.2$ & $<95.5\pm34.9$ \\
9 & $0.71\pm0.10$ & $0.22\pm0.06^*$ & $163.3\pm56.9$ & $3.0\pm1.2$ & $19.6\pm10.5$ & $54.6\pm28.4$ \\
10 & $6.20\pm0.57$ & $0.32\pm0.06^*$ & $383.0\pm87.2$ & $26.0\pm9.6$ & $81.5\pm36.3$ & $14.7\pm6.4$ \\
11 & $7.20\pm0.59$ & $0.40\pm0.04^*$ & $995.3\pm144.2$ & $30.3\pm11.1$ & $60.2\pm23.9$ & $32.9\pm13.0$ \\
12 & $10.41\pm0.82$ & $0.40\pm0.06^*$ & $646.0\pm115.6$ & $43.7\pm16.0$ & $86.8\pm36.0$ & $14.8\pm6.0$ \\
13 & $3.57\pm0.30$ & $<0.84^{**}$ & $<1186.6\pm319.3$ & $15.0\pm5.5$ & $>6.8\pm3.3$ & $<79.1\pm36.0$ \\
14 & $4.15\pm0.41$ & $<0.84^{**}$ & $<682.9\pm191.7$ & $17.5\pm6.5$ & $>7.9\pm3.8$ & $<39.1\pm18.2$ \\
15 & $22.64\pm1.87$ & $<0.84^{**}$ & $<2118.7\pm375.3$ & $95.1\pm34.9$ & $>42.9\pm17.9$ & $<22.3\pm9.1$ \\
16 & $216.18\pm4.17$ & $0.48\pm0.01^*$ & $902.6\pm19.3$ & $908.0\pm324.7$ & $1259.3\pm452.0$ & $1.0\pm0.4$ \\
17 & $101.63\pm3.43$ & $0.33\pm0.01^*$ & $1171.5\pm103.4$ & $426.8\pm153.1$ & $1276.8\pm462.0$ & $2.7\pm1.0$ \\
18 & $66.91\pm2.76$ & $0.44\pm0.02^*$ & $498.5\pm37.1$ & $281.0\pm101.0$ & $463.7\pm169.8$ & $1.8\pm0.7$ \\
19 & $1.28\pm0.15$ & $<0.84^{**}$ & $<795.9\pm280.7$ & $5.4\pm2.0$ & $>2.4\pm1.2$ & $<148.0\pm76.2$ \\
20 & $3.32\pm0.31$ & $0.33\pm0.05^*$ & $1357.4\pm297.0$ & $14.0\pm5.2$ & $40.0\pm17.3$ & $97.3\pm41.8$ \\
21 & $1.32\pm0.12$ & $0.41\pm0.09^*$ & $243.0\pm74.9$ & $5.6\pm2.1$ & $10.4\pm5.0$ & $43.8\pm21.0$ \\
22 & $107.24\pm8.92$ & $0.71\pm0.14$ & $1225.4\pm253.0$ & $450.4\pm165.2$ & $282.2\pm128.7$ & $2.7\pm1.1$ \\
23 & $112.54\pm9.15$ & $0.67\pm0.12$ & $1578.5\pm310.5$ & $472.7\pm173.1$ & $336.6\pm151.3$ & $3.3\pm1.4$ \\
24 & $1.29\pm0.12$ & $<0.84^{**}$ & $<1355.9\pm424.4$ & $5.4\pm2.0$ & $>2.4\pm1.2$ & $<250.7\pm121.3$ \\
\hline
\end{tabular}
\tablefoot{The table does not include clump no. 24 because it includes emission from sidelobes of other stronger clumps. \tablefoottext{*}{Minor axis is unresolved.} \tablefoottext{**}{Both axes are unresolved.}}
\end{table*}

We determined the radius $R$, the $^{12}$CO luminosity $L_{^{12}CO}$, the virial mass $M_{vir}$ and molecular mass derived from CO luminosity (luminous mass) $M^{CO}_{\mathrm{gas}}$ of each clump based on the CPROPS results. The physical properties are corrected for sensitivity and resolution bias. The corrections are applied to the second moments $\sigma_r$ and $\sigma_v$ as described in \cite{rosolowsky2006}. The correction for sensitivity is done as part of the CPROPS routine and we correct for resolution bias separately. We do not extrapolate the radii for clumps no. 7, 8, 16, 17 and 18 (which are characterized using the manual method described in Sect. \ref{sec:cprops}). To correct for resolution bias, we deconvolved the beam size and the width of a spectral channel from $\sigma_r$ and $\sigma_v$, respectively, using Equations 9 and 10 from \cite{rosolowsky2006} 

Out of the 24 clumps, only 3 have second moments $\sigma_r$ along both principal axes that are larger than the geometric mean of the second moments of the beam $\sigma_{beam}$ (i.e., are resolved). For the rest, 11 clumps have a minor axis $\sigma_r$ smaller than $\sigma_{beam}$ and the rest have both axes smaller than $\sigma_{beam}$. We calculated the radii of the clumps taking this into account. When the minor axis is smaller than $\sigma_{beam}$, instead of using Equation 9 of \cite{rosolowsky2006} ,we calculated $\sigma_r$ using:
\begin{equation}
	\sigma_r = \sqrt{\sigma_{beam}(\sigma_{maj}(0K) - \sigma_{beam})},\label{eq:nonresolvedradii}
\end{equation}
where $\sigma_{maj}(0K)$ is the extrapolated second moment of the major axis. If both axes are smaller than the beam, the clump is unresolved and we adopted $\sigma_r=\sigma_{beam}$. This implies that these clumps have an upper limit to the radius of 0.84 pc.

We used the radii $R$ and luminosities $L_{^{12}CO}$ calculated by CPROPS \citep[][]{rosolowsky2006}, assuming a distance to 30 Dor of $D=50$ kpc \citep{Pietrzynski2013}. 
The virial mass is calculated using Eq. 3 of \cite{maclaren1988}, which assumes that clumps have a spherical shape with a density profile $\rho(r) \propto r^{-1}$:
\begin{equation}
    M_{vir}=190(\Delta v)^2R, \label{eq:virialmass}
\end{equation}
where $R$ is the radius of the clump in pc and $\Delta v$ is its FWHM in km s$^{-1}$and gives the mass in $M_{\odot}$.

We also calculated the H$_2$ gas mass traced by $^{12}$CO luminosity, $M^{CO}_{\mathrm{gas}}$, in M$_{\odot}$, as:
\begin{equation}
    M^{CO}_{\mathrm{gas}}=\alpha_{^{12}CO}L_{^{12}CO(1-0)}\label{eq:mco}
\end{equation}
where $\alpha_{^{12}CO}$ is the CO-to-H$_2$ conversion factor in M$_{\odot}$ (K km s$^{-1}$ pc$^2$)$^{-1}$ and $L_{^{12}CO(1-0)}$ is the $^{12}$CO $J=1-0$ luminosity of each clump. We transform the $^{12}$CO $J=3-2$ luminosities of our clumps into $^{12}$CO $J=1-0$ luminosities, using a line ratio $R_{\frac{3-2}{1-0}}\sim2$, as determined by \cite{johansson1998} for the 30Dor-10 cloud. 
We used $\alpha_{^{12}CO}=8.4\pm3.0$  M$_{\odot}$ pc$^{-2}$ (K km s$^{-1}$)$^{-1}$, obtained toward the 30Dor-10 cloud by \cite{indebetouw2013}, at 0.6 pc resolution. This value is almost double the canonical $\alpha_{^{12}CO}$ for the Milky Way \citep[4.3 M$_{\odot}$ pc$^{-2}$ (K km s$^{-1}$)$^{-1}$,][]{bolatto2013}. The CO-to-H$_2$ conversion factor is known to vary significantly on small scales ($\sim1$ pc), as shown by excitation analyses toward Molecular Clouds with diverse physical conditions \citep[][]{Goldsmith2008taurus,Kohno2024Xcovar}. This does not affect the obtained masses considerably, given the unresolved nature of the majority of our clumps.

We calculated the H$_2$ surface density $\Sigma_{\text{H}_2}$ in M$_{\odot}$ pc$^{-2}$ using the resulting mass from $^{12}$CO luminosity and the radius of each clump. 

\begin{figure}[ht]
    \centering
    \includegraphics[width=0.49\textwidth]{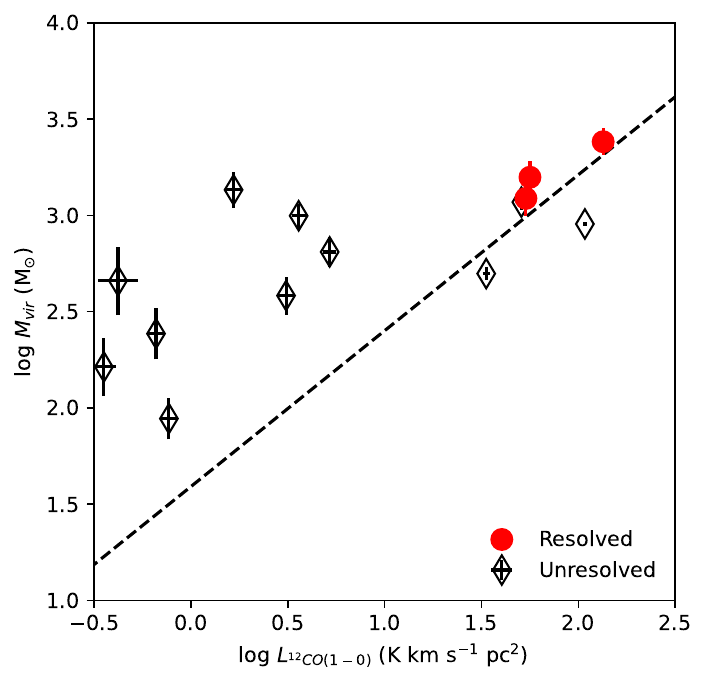}
    \caption{Relationship between the virial mass $M_{vir}$ and the $^{12}$CO $J=1-0$ luminosity $L_{^{12}CO(1-0)}$. The $L_{^{12}CO}$ luminosities are transformed to $L_{^{12}CO(1-0)}$ using $R_{\frac{3-2}{1-0}}\sim2$ \citep{johansson1998}. Filled red circles correspond to completely resolved clumps, whereas empty red circles correspond to clumps where the minor axis is unresolved. The dashed line represents the relationship found for Molecular Clouds in the first Galactic quadrant by \cite{solomon1987}, $M_{vir}=39(L_{^{12}CO(1-0)})^{0.81}$.}
    \label{fig:larsonlaws30dor}
\end{figure}

Table \ref{tab:propclouds} presents the physical properties obtained for all clumps.  We also add for reference the ratio between the virial mass and the luminous mass using the LMC conversion factor, ${M_{vir}}/{M^{CO}_{\mathrm{gas}}}$. The clumps have radii between 0.1 and 0.9 pc, luminosities between 1 and 269 K km s$^{-1}$ pc$^{2}$, virial masses between 88 and 2414 M$_{\odot}$ and luminous masses $M^{CO}_{\mathrm{gas}}$ between 3 and 1128 M$_{\odot}$. $M_{vir}$ tend to be larger than $M^{CO}_{\mathrm{gas}}$ except clump no. 16, in which the ratio between both masses is 1. As the majority of the clumps are not resolved, their virial masses are an upper limit, so their mass ratios are an upper limit as well. For the completely resolved clumps (no. 2, 22, and 23), the ratio is between 2 and 3.3. These results are similar to the ratios found at lower resolution by \cite{kalari2018}. This difference could be explained if the conversion factor in this region is a factor of $2-3$ different from the one we assumed, $\alpha_{^{12}CO}=8.4\pm3.0$  M$_{\odot}$ pc$^{-2}$ (K km s$^{-1}$)$^{-1}$. If we assume $M_{vir}\approx M_{H_2}$, the conversion factor for this region would be between $\alpha_{^{12}CO}=16.8$ and 27.7 M$_{\odot}$ pc$^{-2}$ (K km s$^{-1}$)$^{-1}$. These $\alpha_{^{12}CO}$ values are between 4 to 6 times larger than the canonical galactic conversion value of $\alpha_{^{12}CO}=4.3$  M$_{\odot}$ pc$^{-2}$ (K km s$^{-1}$)$^{-1}$. A larger $\alpha_{^{12}CO}$ is consistent with the large fraction of CO-dark gas in this region \citep{Chevance2020}. The difference between $M^{CO}_{\mathrm{gas}}$ and $M_{vir}$ can also be the result of the virial mass tracing the external pressure suffered by the clumps as well as the total gas mass. We discuss the virial mass in further detail in Sect. \ref{30Dor:comparisonsCO} and \ref{sec:gasmassdiscussion}.

We plot $M_{vir}$ v/s $L_{^{12}CO(1-0)}$ for the molecular clumps which are completely resolved and clumps which have their minor axis unresolved in Figure \ref{fig:larsonlaws30dor}, together with the relation between these properties found for Molecular Clouds in the first quadrant of the Milky Way by \cite{solomon1987}, $M_{vir}=39(L_{^{12}CO(1-0)})^{0.81}$. The $L_{^{12}CO}$ luminosities are transformed to $L_{^{12}CO(1-0)}$ using $R_{\frac{3-2}{1-0}}\sim2$ \citep{johansson1998}. The clumps with luminosities $L_{^{12}CO(1-0)}>10$ K km s$^{-1}$ pc$^2$ (no. 2, 16, 17, 18, 22 and 23) follow the relation between $M_{vir}$ and $L_{^{12}CO(1-0)}$, whereas the clumps with $L_{^{12}CO(1-0)}<10$ K km s$^{-1}$ pc$^2$ have a larger virial mass than in \cite{solomon1987} relation. The latter clumps are unresolved in one of their axis, so the virial mass in these cases is overestimated as we truncate their size in the minor axis to the size of the beam. Higher resolution observations will be able to resolve these clumps and may lower these points toward the relationship found for Galactic Molecular Clouds \citep{solomon1987}.

\subsection{Physical properties derived from $^{12}$CO  and $^{13}$CO molecular emission\label{sec:columndensities}}

For the first time, we have simultaneous $^{12}$CO and $^{13}$CO molecular line emissions at the same resolution and line transition towards the molecular gas near R136 in 30 Dor. We calculated the mass of the clumps using both molecules assuming Local Thermodynamic Equilibrium (LTE), as an alternative to assuming an $\alpha_{CO}$ conversion factor. To obtain the column density of H$_2$ molecules $N(H_2)$ for the peak positions of the clouds under LTE assumption, we used the equations specific for $^{12}$CO $J=3-2$ and $^{13}$CO $J=3-2$ molecular line emissions given in \cite{celispena2019}.  

First, we assumed that the excitation temperature for $^{13}$CO is the same than for $^{12}$CO, and that the $^{12}$CO emission is optically thick. Thus we obtained the excitation temperature $T_{ex}(^{12}CO)=T_{ex}$ for the $^{12}$CO molecular transition in K using:

\begin{equation}
    T_{ex} = \frac{16.59}{\ln\left(1+\frac{16.59}{T_{peak}+0.036}\right)}\label{eq:tex}
\end{equation}
where $T_{peak}$ is the $^{12}$CO peak temperature of the clump in K. We used the $T_{peak}$ values in Table \ref{tab:cpropsresults} for each clump where we detect $^{13}$CO emission. We note that Eq. \ref{eq:tex} assumes a beam filling factor $f\approx 1$, which might not be true for unresolved clumps. We discuss the effect of this assumption further in Sect. \ref{sec:lineratiocomparison}.

We confirmed that the $^{13}$CO line is optically thin in all clumps by calculating the optical depth $\tau^{13CO}$:

\begin{equation}
    \tau^{13CO} = - \ln \left( 1-\frac{0.063\,T_{peak}^{^{13}CO}}{(e^{\frac{15.85}{T_{ex}}}-1)^{-1}-0.003}\right),
\end{equation}
where $T_{peak}^{^{13}CO}$ is the $^{13}$CO peak temperature for the clump in K (obtained from CPROPS, Appendix \ref{appendix:cpropsmolecules}). The resulting $\tau^{13CO}$ are all lower than 1, between 0.1 and 0.4 (Table \ref{tab:textauandN}).

We calculated the $^{13}$CO column density $N(^{13}CO)$ in cm$^{-2}$ using the optically thin approximation:
\begin{equation}
    N(^{13}CO) = 8.28\times10^{13} e^{\frac{15.85}{T_{ex}}} \frac{T_{ex}+0.88}{1-e^{\frac{-15.85}{T_{ex}}}} \frac{ I_{^{13}CO}}{J(T_{ex})-J(T_{BG})},
\end{equation}
where $I_{^{13}CO}$ is the integrated line intensity at the peak temperature position in K km s$^{-1}$, obtained from Table \ref{tab:intensities}, $T_{BG}$ is the cosmic microwave background (CMB) radiation temperature $T_{BG}=2.73$ K and
\begin{equation}
    J(T) = \frac{h\nu/k}{e^{h\nu/kT}-1},
\end{equation}
where, for $\nu=330.588$ GHz (the rest frequency of the $^{13}$CO $J=3-2$ line), $h\nu/k=15.85$ K.

We obtained the column density of H$_2$ molecules $N(H_2)$ in cm$^{-2}$ as $N(H_2)=[H_2/^{13}CO]N(^{13}CO)$, where $[H_2/^{13}CO]$ is the abundance ratio between H$_2$ and $^{13}$CO, which we assumed to be $1.8\times10^6$ \citep{garay2002, heikkila1999}. 

Finally, we derive the gas mass of the clumps $M^{LTE}_{gas}$ using:
\begin{equation}
M^{LTE}_{gas} = \mu m_{H}D^2 \Omega_{beam}N(H_2),
\end{equation}
where $\mu$ is the mean molecular weight of H$_2$, equal to 2.72 to include the contribution of Helium to the total mass of the clump, $m_{H}$ is the mass of the Hydrogen atom in gr, $D$ is the distance to the clump in cm and $\Omega_{beam}$ is the solid angle covered by the beam in sr. 

Additionally, we obtained the $^{13}$CO luminosity $L_{^{13}CO}$ applying the mask from the $^{12}$CO clump determination to the $^{13}$CO data, following the same luminosity definition as CPROPS \citep[see][for more details]{rosolowsky2006} and again assuming a distance to 30 Dor of 50 kpc \citep{Pietrzynski2013}. We do not correct for sensitivity bias as done for the $^{12}$CO luminosity. The obtained $T_{ex}$, $\tau^{13CO}$,  $N(^{13}CO)$, $N(H_2)$, $L_{^{13}CO}$ and $M^{LTE}_{gas}$ are in Table \ref{tab:textauandN}. We did not include clump no. 9 as it has a low $S/N\sim 3$ in  $^{13}$CO.

The clumps have excitation temperatures $T_{ex}$ ranging from 5.38 to 33.39. The highest temperatures are found in the brightest clumps, no. 2, 16, 17, 18, 22, and 23, belonging to the northwest-southeast structure. These clumps were also detected in HCO$^{+}$, CS and/or HCN, so they have a high density ($n\sim10^{6}$ cm$^{-3}$) and thus, one could assume that the derived $T_{ex}\approx T_K$. We compare $T_{ex}$ in this region with other regions in the LMC in Sect. \ref{sec:lineratiocomparison}.

The clumps' peak column densities range between $4.2-23.7\times10^{21}$ cm$^{-2}$. Those clumps with column densities exceeding $10^{22}$ cm$^{-2}$, clumps no. 2, 8, 16, 17, and 18, are those with the strongest CO emission. The total gas masses $M^{LTE}_{gas}$ range between 203 and 1148 M$_{\odot}$. For the resolved clumps (no. 2, 22, and 23), these masses are between $0.6-1$ times the mass derived from CO luminosity $M^{CO}_{\mathrm{gas}}$ from Table \ref{tab:propclouds}. For the rest of the clumps, these masses are between $1.5-10.5$ times larger than $M^{CO}_{\mathrm{gas}}$, except for clump no. 16, where $M^{LTE}_{gas}\approx 0.5M^{CO}_{\mathrm{gas}}$. The fact that $M^{LTE}_{gas}> M^{CO}_{\mathrm{gas}}$ for unresolved clumps suggests that the assumption that $^{12}$CO and $^{13}$CO have a beam filling factor $f\approx 1$ is incorrect for these clumps. This increases the estimated $T_{ex}$ and $N(^{13}CO)$. The difference between $M^{CO}_{\mathrm{gas}}$ and $M^{LTE}_{gas}$ also suggests that there are local variations in the $\alpha_{CO}$ factor within the clumps, as mentioned in Sect. \ref{sec:physprops} \citep[e.g.,][]{Kohno2024Xcovar}. Given that only three clumps are resolved in this sample, a further investigation onto $\alpha_{CO}$ is beyond the scope of this work.

\begin{table*}[htb]
\caption{Physical properties of clumps derived through an LTE analysis of $^{12}$CO and $^{13}$CO emission.\label{tab:textauandN}}
\centering
\begin{tabular}{ccccccc}
\hline\hline
ID & $T_{ex}$  & $\tau^{^{13}CO}$ & $N(^{13}CO)$ & $N(H_2)$ &  $L_{^{13}CO}$ & $M^{LTE}_{gas}$  \\
 & (K) &  &  ($\times10^{15}$ cm$^{-2}$) & ($\times10^{21}$ cm$^{-2}$) & (K \kms pc$^2$)  & (M$_{\odot}$) \\
\hline
2 & $33.39\pm0.06$ & $0.38\pm0.00$ & $13.2\pm1.3$ & $23.7\pm2.4$ & $58.43\pm5.86$ & $1148\pm115$ \\
8 & $5.38\pm0.33$ & $0.42\pm0.08$ & $7.1\pm1.4$ & $12.7\pm2.5$ & $0.44\pm0.11$ & $616\pm123$ \\
10 & $7.11\pm0.19$ & $0.29\pm0.03$ & $3.1\pm0.4$ & $5.7\pm0.8$ & $0.88\pm0.15$ & $274\pm38$ \\
11 & $6.38\pm0.23$ & $0.12\pm0.04$ & $3.1\pm0.6$ & $5.5\pm1.1$ & $0.37\pm0.12$ & $266\pm54$ \\
12 & $7.61\pm0.17$ & $0.15\pm0.03$ & $2.8\pm0.4$ & $5.0\pm0.7$ & $0.78\pm0.16$ & $243\pm33$ \\
14 & $6.76\pm0.21$ & $0.30\pm0.04$ & $3.3\pm0.5$ & $5.9\pm0.8$ & $0.69\pm0.13$ & $285\pm41$ \\
15 & $10.45\pm0.12$ & $0.17\pm0.01$ & $2.3\pm0.3$ & $4.2\pm0.5$ & $2.39\pm0.30$ & $203\pm23$ \\
16 & $32.76\pm0.06$ & $0.23\pm0.00$ & $5.7\pm0.6$ & $10.2\pm1.0$ & $23.96\pm2.41$ & $494\pm50$ \\
17 & $16.13\pm0.08$ & $0.24\pm0.01$ & $7.1\pm0.7$ & $12.8\pm1.3$ & $9.77\pm1.01$ & $619\pm62$ \\
18 & $15.07\pm0.09$ & $0.15\pm0.01$ & $6.2\pm0.6$ & $11.2\pm1.1$ & $7.19\pm0.76$ & $543\pm55$ \\
21 & $4.86\pm0.43$ & $0.27\pm0.10$ & $4.0\pm1.5$ & $7.3\pm2.8$ & $0.15\pm0.07$ & $352\pm134$ \\
22 & $20.37\pm0.07$ & $0.20\pm0.00$ & $3.3\pm0.3$ & $6.0\pm0.6$ & $14.62\pm1.49$ & $290\pm29$ \\
23 & $18.60\pm0.08$ & $0.28\pm0.01$ & $4.5\pm0.5$ & $8.1\pm0.8$ & $17.69\pm1.79$ & $390\pm39$ \\\hline         
\end{tabular}

\end{table*}

\subsection{Integrated line intensities for different molecular species \label{sec:mollinesint}}

We compared the different molecules using their line intensities and their luminosities. We determined the integrated line intensity $I=\int T_v\, dv$ using the spectra at the peak emission position for each clump in each molecule. We first fit a Gaussian profile to the spectra using the \texttt{astropy.models.Gaussian1D} module. Then, we integrated the best fit Gaussian profile of each molecule to obtain the velocity integrated line intensity, $I=\int T_v dv$. For clumps no. 8 and no. 18, which are the clumps manually characterized in Sect. \ref{sec:cpropssecondstep}, the detected molecules were fit with two Gaussian components, the second of which corresponds to emission from clumps no. 7 in the case of clump no. 8, and 17 in the case of clump no. 18, located in the same lines of sight. In these cases, we selected the Gaussian component that has the central velocity closest to the velocity of the clump in Table \ref{tab:cpropsresults}. Emission from clumps no. 4, 6 and 24 cannot be easily disentangled from other clumps and are weak ($S/N\sim5$), therefore we do not include them in this analysis.

Table \ref{tab:intensities} shows the integrated line intensities for all the clumps identified in the mapped region. The obtained spectra for each molecule in each clump, together with the best fit Gaussian models, are plotted in Appendix \ref{appendix:molspectra}. 
In general, the strongest intensities $I$ for in all molecular species are found in the clumps which belong to the northwest-southeast diagonal (clumps 2, 16, 17, 18, 22, and 23).

\begin{table*}[ht]
\caption{\label{tab:intensities}Integrated line intensities $I=\int T_v dv$ for each of the clumps. }
\centering
\begin{tabular}{@{}llllll@{}}
\hline\hline
ID & $I_{^{12}CO}$ & $I_{^{13}CO}$ & $I_{CS}$& $I_{HCO^{+}}$  & $I_{HCN}$  \\ 
   & (K km s$^{-1}$)           & (K km s$^{-1}$)    & (K km s$^{-1}$)      & (K km s$^{-1}$)         & (K km s$^{-1}$)       \\\hline
1  &     $1.98 \pm 0.20$ &     - &    - &     - &    - \\
2  &  $125.12 \pm 12.51$ &  $28.46 \pm 2.85$ &  $1.91 \pm 0.20$ &  $14.20 \pm 1.42$ &  $3.13 \pm 0.32$ \\
3  &     $1.37 \pm 0.15$ &     - &    - &     - &    - \\
5  &     $1.13 \pm 0.13$ &     - &    - &     - &    - \\
8  &     $1.68 \pm 0.31$ &   $0.49 \pm 0.10$ &    - &   $0.28 \pm 0.10$ &    - \\
9  &     $0.69 \pm 0.09$ &   $0.18 \pm 0.07$ &   - &   $0.07 \pm 0.06$ &   - \\
10 &     $5.40 \pm 0.54$ &   $0.85 \pm 0.12$ &  $0.18 \pm 0.05$ &   $1.02 \pm 0.12$ &  $0.29 \pm 0.06$ \\
11 &     $5.57 \pm 0.56$ &   $0.54 \pm 0.11$ &    - &   $0.35 \pm 0.08$ &    - \\
12 &     $8.23 \pm 0.83$ &   $0.95 \pm 0.13$ &    - &   $0.73 \pm 0.10$ &    - \\
13 &     $3.06 \pm 0.32$ &     - &    - &     - &    - \\
14 &     $3.92 \pm 0.40$ &   $0.73 \pm 0.10$ &    - &   $0.19 \pm 0.06$ &    - \\
15 &    $16.82 \pm 1.68$ &   $1.87 \pm 0.21$ &    - &   $2.04 \pm 0.22$ &    - \\
16 &    $74.78 \pm 7.48$ &  $12.26 \pm 1.23$ &  $0.85 \pm 0.10$ &   $7.36 \pm 0.74$ &  $1.38 \pm 0.15$ \\
17 &    $29.53 \pm 5.70$ &  $5.57 \pm 0.79$ &  $0.48 \pm 0.11$ &   $5.96 \pm 0.60$ &  $1.29 \pm 0.15$ \\
18 &    $26.67 \pm 4.90$ &   $3.53 \pm 0.58$ &  $0.13 \pm 0.05$ &     - &    - \\
19 &     $1.99 \pm 0.21$ &     - &    - &     - &    - \\
20 &     $2.42 \pm 0.26$ &     - &    - &     - &    - \\
21 &     $1.06 \pm 0.12$ &   $0.18 \pm 0.07$ &    - &   $0.18 \pm 0.06$ &    - \\
22 &    $44.57 \pm 4.46$ &   $6.31 \pm 0.64$ &  $0.11 \pm 0.06$ &   $1.22 \pm 0.14$ &  $0.10 \pm 0.06$ \\
23 &    $46.36 \pm 4.64$ &   $7.98 \pm 0.80$ &  $0.39 \pm 0.07$ &   $2.26 \pm 0.24$ &  $0.36 \pm 0.07$ \\
\hline
\end{tabular}

\end{table*}

\begin{table*}[ht]
\caption{\label{tab:intlineratios}Intensity line ratios calculated using the results shown in Table \ref{tab:intensities}. }
\centering
\begin{tabular}{llllll}
\hline\hline
ID & $\frac{I_{^{12}CO}}{I_{^{13}CO}}$ &  $\frac{I_{CS}}{I_{^{13}CO}}$ & $\frac{I_{HCO^{+}}}{I_{^{13}CO}}$ &  $\frac{I_{HCN}}{I_{^{13}CO}}$ & $\frac{I_{HCO^{+})}}{I_{HCN}}$ \\ \hline
2  &   $4.40 \pm 0.62$ &  $0.07 \pm 0.01$ &  $0.50 \pm 0.07$ &  $0.11 \pm 0.02$ &   $4.54 \pm 0.65$ \\
8  &   $3.41 \pm 0.95$ &    - &  $0.69 \pm 0.23$ &    - &     - \\
10 &   $6.36 \pm 0.98$ &  $0.22 \pm 0.08$ &  $1.20 \pm 0.21$ &  $0.34 \pm 0.09$ &   $3.51 \pm 0.69$ \\
11 &  $10.30 \pm 1.80$ &    - &  $0.64 \pm 0.21$ &    - &     - \\
12 &   $9.66 \pm 1.48$ &    - &  $0.77 \pm 0.15$ &    - &     - \\
14 &   $5.37 \pm 0.84$ &    - &  $0.26 \pm 0.10$ &    - &     - \\
15 &   $8.98 \pm 1.31$ &    - &  $1.09 \pm 0.17$ &    $0.11\pm0.04$ &     $9.63\pm2.07$ \\
16 &   $6.10 \pm 0.86$ &  $0.07 \pm 0.01$ &  $0.60 \pm 0.09$ &  $0.11 \pm 0.02$ &   $5.33 \pm 0.77$ \\
17 &   $5.30 \pm 1.27$ &  $0.09 \pm 0.02$ &  $1.07 \pm 0.19$ &  $0.23 \pm 0.04$ &   $4.62 \pm 0.68$ \\
18 &   $7.55 \pm 1.85$ &  $0.04 \pm 0.01$ &    - &    - &     - \\
22 &   $7.06 \pm 1.00$ &  - &  $0.19 \pm 0.03$ &  - &  - \\
23 &   $5.81 \pm 0.82$ &  $0.05 \pm 0.01$ &  $0.28 \pm 0.04$ &  $0.04 \pm 0.01$ &   $6.31 \pm 1.14$ \\
\hline
\end{tabular}

\end{table*}

We determined the integrated intensity line ratios with respect to the $^{13}$CO line $\frac{I_{^{12}CO}}{I_{^{13}CO}}$, $\frac{I_{CS}}{I_{^{13}CO}}$,  $\frac{I_{HCO^{+}}}{I_{^{13}CO}}$, and  $\frac{I_{HCN}}{I_{^{13}CO}}$ based on Table \ref{tab:intensities}. 
The resulting line ratios are shown in Table \ref{tab:intlineratios}, where we also include in the last column the $\frac{I_{HCO^{+}}}{I_{HCN}}$ line ratio. We did not calculate the line ratios for clumps no. 9 and 21 because their $^{13}$CO and HCO$^{+}$ detections have a low $S/N$ ratio (between $3-4$), which generates a higher uncertainty for the ratio. We also do not report the line ratios using CS, HCO$^{+}$, and HCN detections for clump no. 22 because these detections also have a low $S/N$ ($\sim3$).

\subsection{Physical properties of the dust clumps\label{sec:propsdust}}

\subsubsection{Free-free emission near R136\label{sec:freefree}}

The flux measurements from the ALMA 0.88 mm continuum image (Table \ref{tab:contflux}) are in part due to dust emission, but might also contain free-free (Bremsstrahlung) emission from ionized gas and synchrotron emission from relativistic particles. We are interested in the dust continuum emission, therefore we need to calculate and remove the other contributions to the measured flux. Previous works show that synchrotron emission is negligible in 30 Dor \citep{brunetti2019, guzman2010}. However, we expect the ALMA 0.88 mm continuum image to have an important contribution from free-free emission, as the clumps are within an \ion{H}{II} region. 

\begin{figure}[ht]
	\centering
	\includegraphics[width=0.45\textwidth]{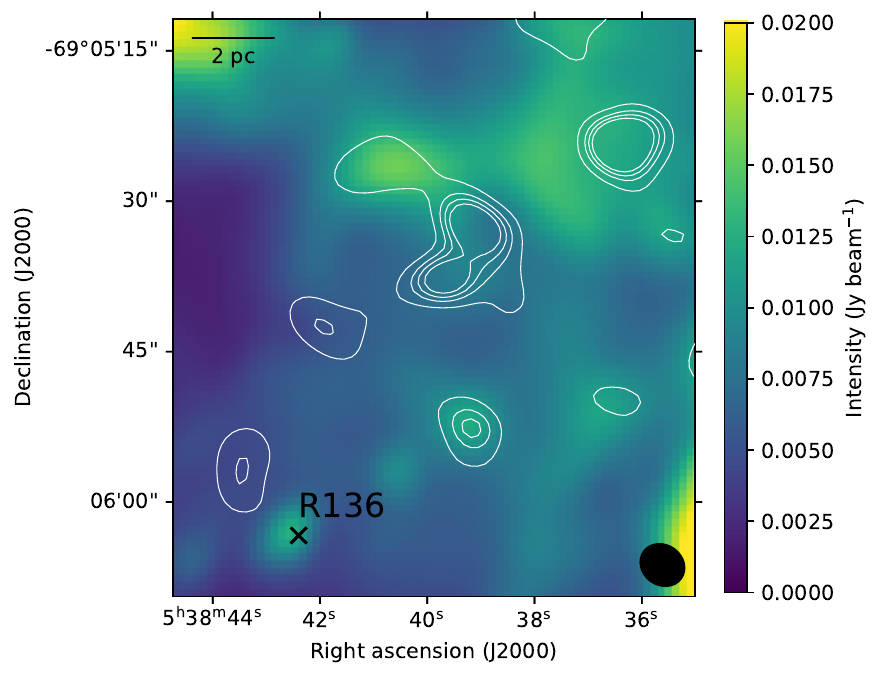}
	\caption{Free-free emission image generated from Br$\gamma$ emission. The white dashed contours are the 0.88 mm continuum contours from Fig. \ref{fig:continuumALMA}. The scale-bar on the top left represents a 2 pc length. The black ellipse shows the resolution of the free-free image after convolution. \label{fig:freefreecomp} }
\end{figure}

We determined the free-free emission in the vicinity of R136 using a Br$\gamma$ emission image obtained with the 1.5 m telescope in Cerro Tololo Observatory (M. Rubio, priv. comm.). The image has a resolution of $1.2\arcsec$ and intensity units of erg cm$^{-2}$ s$^{-1}$ sr$^{-1}$. We first transformed the Br$\gamma$ emission image into an H$\alpha$ image using a ratio between H$\alpha$ and Br$\gamma$ intensities of 101.78, which corresponds to the ratio for a typical \ion{H}{II} region with an electronic temperature $T_e=10^4$ and electron density of 100 cm$^{-3}$ \citep{osterbrock2006}. Then, we transformed the H$\alpha$ intensity $I_{\alpha}$ in erg cm$^{-2}$ s$^{-1}$ sr$^{-1}$ to a free-free emission image $I_{\nu}^{ff}$ in mJy sr$^{-1}$, using:
\begin{equation}
    I_{\nu}^{ff} = 1.16 \left(1+\frac{n(\text{He}^+)}{n(\text{H}^+)}\right) \left(\frac{T_e}{10^4}\right)^{0.62} \nu^{-0.1}\left( \frac{I_{\alpha}}{10^{-12}}\right)
    \label{eq:halpha}
\end{equation}
\citep[derived from][]{hunt2004} where $\nu$ is the objective frequency in GHz (in this case, 338.5 GHz), $T_e$ is the electronic temperature in K and $n(He^{+})/n(H^{+})$ is the number density ratio between He and H ions. We used $T_e=10^4$ K and $n(He^{+})/n(H^{+})\sim0.08$, typical values estimated for low metallicity sources like the Magellanic Clouds \citep{hunt2004}. We convolved the image to the same resolution of $4.7\times3.9\arcsec$ as the ALMA 0.88 mm continuum image and transformed the image units from mJy sr$^{-1}$ to m\Jyb by multiplying each pixel by the area of the beam in sr. The resulting free-free emission image is shown in Fig. \ref{fig:freefreecomp} and has an rms of 4.5 m\Jyb. 

We determined the free-free emission towards each clump through aperture photometry. 
We used the same apertures from Sect. \ref{sec:contradii}. The obtained free-free fluxes, $S_{ff, Br\gamma}$, derived from Br$\gamma$ are listed in Table \ref{tab:contflux}. 
These values represent between 50 and 85\% of the total 0.88 mm flux measured in the ALMA images, confirming that the free-free contribution to the continuum is important, as expected toward an \ion{H}{II} region. Clump C presents the largest free-free contribution with respect to the total 0.88 mm emission. In Figure \ref{fig:freefreecomp}, there is a local free-free peak at the location of clump C, which explains why the free-free emission contribution to its total flux is so important.

\begin{table}[ht]
\caption{\label{tab:contflux} Free-free emission fluxes, resulting dust emission fluxes $S_{dust}$ and total gas masses $M_{gas}^{dust}$ for each clump. }
\centering
\begin{tabular}{cccc}
\hline\hline
ID  & $S_{ff, Br\gamma}$ (mJy) & $S_{dust}$ (mJy) & $M_{gas}^{dust}$ \\
 & (mJy) &  (mJy) & (\Msun) \\ \hline
A & $32.2\pm9.6$ & $29.0\pm17.2$ & $741\pm440$ \\
B & $37.5\pm11.7$ & $11.6\pm21.1$ & $296\pm539$ \\
C & $16.5\pm7.6$ & $2.9\pm13.7$ & $74\pm350$ \\
D & $22.1\pm10.4$ & $10.7\pm18.7$ & $273\pm478$ \\
E & $7.3\pm8.2$ & $10.4\pm14.8$ & $266\pm378$ \\
F & $4.9\pm8.4$ & $13.8\pm15.1$ & $353\pm386$ \\
\hline
\end{tabular}

\end{table}

\subsubsection{Gas mass from dust emission\label{sec:gasmassfromdust}}

We calculated the total gas masses of the clumps near R136 using the dust fluxes obtained in Sect. \ref{sec:freefree}. This method has been used to obtain the total gas mass of Molecular Clouds in low metallicity environments, where a large fraction of the total molecular mass may not be traced by CO emission \citep[e.g.,][]{rubio2004, bot2010}. The total gas mass of a clump in M$_{\odot}$ was obtained assuming dust emission is optically thin at 0.88 mm:
\begin{equation}
    M_{gas}^{dust} = \frac{S_{dust} D^2}{\kappa(\nu)x_d B_{\nu}(T_d)},\label{eq:dustgasmass}
\end{equation}
where $S_{dust}$ is the dust flux of the clump, 
$D$ is the distance to the clump (in this case, 50 kpc),  $\kappa(\nu)$ is the dust absorption coefficient at frequency $\nu$, $x_d$ is the dust to gas mass ratio and $B_{\nu}(T_d)$ is the Planck law evaluated at a dust temperature $T_d$, in Jy sr$^{-1}$. We calculated the gas masses from dust emission using the dust fluxes $S_{dust}$ from Table \ref{tab:contflux}. We assumed a dust temperature $T_d = 40$ K, based on the dust temperatures obtained for this region in \cite{Tram2021}. We note that the temperatures of individual clumps might be different, but variations within 5 K will not affect our analysis. We also assumed that the dust grain properties in 30 Dor are similar to the ones present in the molecular ring of the Milky Way, such that $\kappa(870\mu m)=1.26\pm0.02$ cm$^2$ g$^{-1}$, found by \cite{bot2010}. We transformed $\kappa(870\mu m)$ to $\kappa(880\mu m)$, using $\kappa(880\mu m) = (880\mu m/870\mu m)^{-\beta} \kappa(870\mu m)$, using a dust emissivity index $\beta=2$, which resulted in $\kappa(880\mu m) = 1.23\pm 0.02$ cm$^2$ g$^{-1}$. We finally assumed that the dust to gas ratio scales linearly with metallicity, and as $Z(LMC)=0.5Z(\odot)$ \citep{rolleston2002}, the dust to gas ratio in the LMC is half the dust to gas ratio in the Solar neighborhood\footnote{$x_{d}(\odot)\sim 0.007$ \citep{Draine2007}}, $x_d(LMC)=0.5x_d(\odot)=3.5\times10^{-3}$. 

The gas masses obtained from Equation \ref{eq:dustgasmass} are in Table \ref{tab:contflux}. All masses are within $10^2$ and $10^3$ M$_{\odot}$, the same orders of magnitude as the virial masses obtained in Sect. \ref{sec:physprops} for the clumps in the northwest-southeast diagonal structure. 

\subsection{Comparison between molecular and continuum emission near R136 \label{sec:coandcontcomparison}}

In this work, the molecular gas in R136 has been studied by the emission of $^{12}$CO and the dust emission by the sub-millimeter 0.88 mm continuum emission, both of which have similar resolution ($4.7\arcsec\times3.9\arcsec$ for the continuum image and $4.6\arcsec\times4.0\arcsec$ for the $^{12}$CO molecular line cube). Therefore, we can make a comparison of some properties derived from both components assuming that the dust continuum emission is associated to the same molecular clumps. 

\subsubsection{Clump sizes in CO and continuum emission\label{sec:extensionscomparison}}

We compared the dust clump sizes from Table \ref{tab:radiicont} with the area covered by the corresponding clumps in $^{12}$CO. We matched the $^{12}$CO clumps in the diagonal structure that dominates the channel maps (Fig. \ref{fig:chanmap-cprops}, between 232.5 and 252.5 \kms) to the dust clumps from Fig. \ref{fig:continuumALMA} via visual inspection. To determine the area of the $^{12}$CO clumps in a similar manner as for the dust clumps, we made velocity integrated images for each clump in the velocity range in which emission is detected. The velocity range used for the integration is  $[v-2\sigma_v,v+2\sigma_v]$ for each clump, where $v$ and $\sigma_v$ correspond to the central velocity and velocity dispersion of the clump given in Table \ref{tab:cpropsresults}. This range contains 95.4\% of the total emission from the clump, assuming it follows a Gaussian distribution in velocity. We determined the rms for the velocity integrated image for each clump and defined the extension of the clump as the area enclosed by the 3$\sigma$ contour. We chose 3$\sigma$ instead of 1.5$\sigma$ (as in Sect. \ref{sec:contradii}) because the integrated images' 1.5$\sigma$ contours cover the sidelobes of $^{12}$CO emission.
The velocity ranges used for integration, the resulting rms of the velocity integrated image and the calculated areas in pc$^{2}$ are in Table \ref{tab:COarea}.

\begin{table}[ht]
	\caption{\label{tab:COarea}Velocity ranges and rms values of the integrated $^{12}$CO images. }
	\centering
	\begin{tabular}{llll}
		\hline\hline
		$^{12}$CO clump  & Dust ID  & $v$ range   & rms (\Jyb    \\
		ID & & (km s$^{-1}$) & \kms)  \\
		\hline
		2            & A              & $246.6 - 253.8$ & 1.3    \\
		16, 17, 18   & B+D    & $238.0 - 249.0$ & 2.0 \\
		15           & C                & $244.4 - 250.6$ & 1.0    \\
		22           & E               & $235.1 - 240.3$ & 1.4  \\
		23           & F              & $232.9 - 238.9$ &  1.0  \\ \hline
	\end{tabular}

\end{table}

In the specific case of the dust sources B and D, we calculated their shared area within the 1.5$\sigma$ contour in the 0.88 mm image, without counting the northern extension that does not have a corresponding $^{12}$CO counterpart. This area corresponds to $A=6.16$ pc$^2$. The $^{12}$CO emission has three different clumps, no. 16, 17, and 18, separated in velocity but in the same line of sight, covering the two dust sources. Thus, we integrated the $^{12}$CO line cube between 238 and 249 \kms and determined the area enclosed in the 3$\sigma$ contour of the resulting image.

\begin{figure*}[ht]
	\begin{center}
	\includegraphics[width=\textwidth]{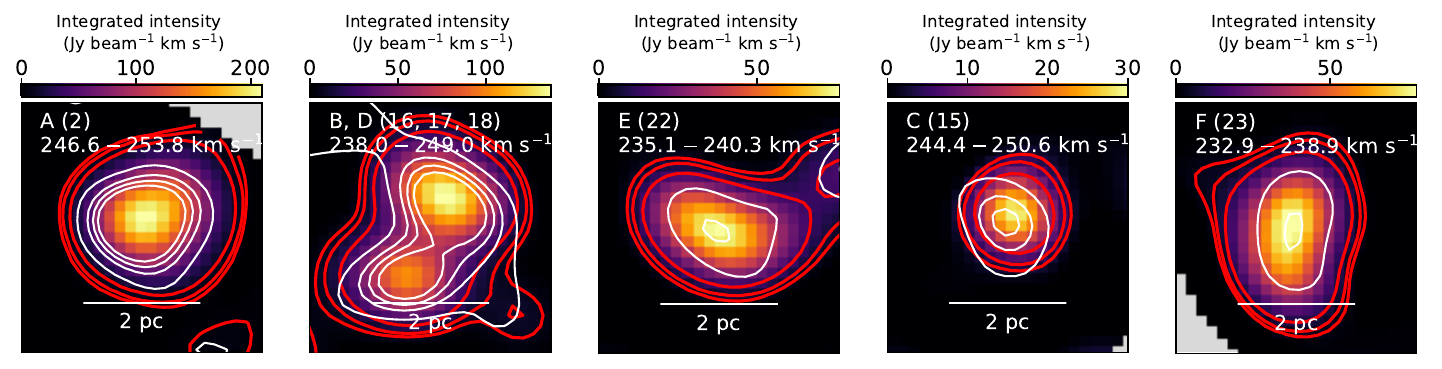}
		\caption{Velocity integrated images of $^{12}$CO emission for clumps no. 2, 15, 16, 17, 18, 22, and 23, together with the corresponding dust continuum contours. The areas of all gas clumps are larger than the areas covered by dust. The red contours correspond to 3, 5 and 10$\sigma$, where $\sigma$ is the rms of each velocity integrated image (Table \ref{tab:COarea}). White contours correspond to the 1.5 and 3$\sigma$ contours of the ALMA 0.88 mm continuum image, as in Figure \ref{fig:continuumALMA}. }
		\label{fig:areacomp}
	\end{center}
\end{figure*}

Figure \ref{fig:areacomp} shows the $^{12}$CO velocity integrated images in the velocity ranges of each clump with the 3, 5 and 10$\sigma$ contours shown in color red. Superimposed to each integrated velocity image we show in color white the ALMA 0.88 mm 1.5 and 3$\sigma$ continuum contours (which correspond to the white contours in Figure \ref{fig:continuumALMA}). In all the identified clumps, the $^{12}$CO velocity integrated images 3$\sigma$ contours cover a larger area than the ALMA 880 dust $\mu$m continuum image 1.5$\sigma$ contour. Also, the calculated areas for the$^{12}$CO clumps in Table \ref{tab:COarea} are larger than the areas from the corresponding 0.88 mm continuum sources in Table \ref{tab:radiicont}. We discuss these results further in Sect. \ref{sec:gasmassdiscussion}.

\subsubsection{Gas masses obtained through CO and dust emission \label{sec:gasandCOmasscomp}}

We also compared the gas mass of the clumps obtained using dust emission and $^{12}$CO line emission from Sect. \ref{sec:physprops}. To perform this comparison we needed to calculate the mass using the same clump areas on all cases. We adopted as sizes the areas obtained using continuum emission, given in Table \ref{tab:radiicont}, as these are the smallest. The gas masses based on dust emission $M_{gas}^{dust}$ were taken from Sect. \ref{sec:gasmassfromdust}. We determined the virial masses associated to the dust clump areas $M_{vir}^{A_{880\mu m}}$ using Equation \ref{eq:virialmass}, adopting the $^{12}$CO velocity FWHM of each clump from Table \ref{tab:propclouds}. To determine the radius, we took the sizes of the dust clumps as given in Table \ref{tab:radiicont} and deconvolved the equivalent radii with the ALMA beam using Equation \ref{eq:dustgasmass}. We also determined the clump masses using $^{12}$CO luminosity $M_{gas}^{CO,A_{880\mu m}}$, assuming the same line ratio and $\alpha_{^{12}CO}$ as in Sect. \ref{sec:physprops}. The resulting $^{12}$CO luminosities and gas masses determined through $^{12}$CO and dust ($M_{CO}^{A_{880\mu m}}$ and $M_{vir}^{A_{880\mu m}}$, respectively) are in Table \ref{tab:Mcocont}. 

In the case of sources B and D together, we calculated $M_{gas}^{dust}$ from the dust flux emission inside the common area, ($A=6.16$ pc$^2$ from Sect. \ref{sec:extensionscomparison}). We subtracted the corresponding free-free emission $S_{ff, Br\gamma}=44.3\pm11.3$ mJy to the measured continuum emission of $S_{880}=80.5\pm9.0$ mJy inside this area. We obtained a total dust flux of $S_{dust}=36.2\pm20.3$ mJy. Using equation \ref{eq:dustgasmass}, with the same $T_d$ and $\epsilon_H$ as in Sect. \ref{sec:gasmassfromdust}, we obtained a gas mass $M_{gas}^{dust}=924\pm520$ M$_{\odot}$ for sources B and D together. The $M_{gas}^{CO,A_{880\mu m}}$ for sources B and D is calculated using the $^{12}$CO luminosity in the chosen area.

\begin{table*}[ht]
\caption{\label{tab:Mcocont}Luminosity and gas masses traced by $^{12}$CO emission, dust emission and their ratios.}
\centering
\begin{tabular}{ccccccccc}
\hline\hline
$^{12}$CO clump & Dust ID & $A_{CO}$  & $L_{^{12}CO^{A_{880\mu m}}}$  & $M_{gas}^{CO,A_{880\mu m}}$ & $M_{vir}^{A_{880\mu m}}$ & $M_{gas}^{dust}$ & $M_{gas}^{CO,A_{880\mu m}}/M_{gas}^{dust}$ & $M_{vir}^{A_{880\mu m}}/M_{gas}^{dust}$  \\
 ID &  & (pc$^2$)  & (K km s$^{-1}$ pc$^2$) & (M$_{\odot}$) & (M$_{\odot}$) & (M$_{\odot}$) & & \\
 \hline
2    & A      & 7.66   & $212.3\pm2.7$    & $892\pm319$       & $2260\pm396$      & $741\pm440$      & $1.2\pm0.9$           & $3.1\pm1.9$          \\
16, 17, 18  & B+D &  9.85  & $214.1\pm6.9$    & $899\pm323$       & $2484\pm589^{\dagger}$ & $924\pm520^{*}$  & $1.0\pm0.8$           & $2.6\pm1.6$         \\
15$^{**}$  & C    & 2.93  & $15.2\pm1.1$     & $64\pm23$         & $2118\pm453$      & $74\pm350$       & $0.9\pm3.8$           & $28.6\pm135.7$           \\
22      & E   & 6.50  & $57.0\pm1.5$     & $239\pm86$        & $1444\pm393$      & $266\pm378$       & $0.9\pm1.3$           & $5.4\pm7.8$            \\
23       & F  & 7.18  & $62.5\pm1.1$     & $263\pm94$        & $1983\pm595$      & $353\pm386$       & $0.8\pm0.9$           & $5.6\pm6.4$    \\ \hline     
\end{tabular}
\tablefoot{
\tablefoottext{$\dagger$}{We used $R_{eq, dc}=1.26\pm0.22$ pc to calculate $M_{vir}$ for clumps no. 16, 17, and 18 together, and we use the FWHM of clump no. 17 as the FWHM because it is the largest FWHM of the three clumps. }
\tablefoottext{*}{Mass of B and D clumps together, obtained from the dust flux in the region shared by both clumps in 0.88 mm continuum emission. }
\tablefoottext{**}{Clump not resolved, so the values are not considered for analysis.}
}

\end{table*}

The gas masses traced by dust emission are similar than the gas masses traced by $^{12}$CO luminosity, even though continuum emission seems to cover a smaller area than $^{12}$CO emission. The ratios $M_{gas}^{CO,A_{880\mu m}}/M_{gas}^{dust}$ for our sample are between 0.8 and 1.2, but in most of them, the uncertainties are large enough that the ratio could be $>1$. 

The ratios $M_{vir}^{A_{880\mu m}}/M_{gas}^{dust}$ for these clumps, on the other hand, are 3.1 and 2.6 for CO clumps no. 2 and the combined emission from no. 16, 17 , and 18 (where the radii in the 0.88 mm continuum image is resolved), with large uncertainties due to the uncertainties in $M_{gas}^{dust}$. The ratio for clumps no. 22 and 23, where the continuum radii is not resolved, is 5.4 and 5.6, respectively. Thus, the observed gas masses obtained from dust, within all uncertainties, are always smaller than the virial mass.

\section{Discussion\label{sec:discussion}}

\subsection{Comparison with molecular emission in previous works\label{30Dor:comparisonsCO}}

Our results confirm the presence of clumpy structures in this region suggested first in in \cite{Rubio2009} using CO $J=2-1$ and CS $J=2-1$ observations. Clump no. 17 coincides in position and velocity with their strongest CO component. Clumps no. 2 and 12 coincide with the other two velocity components.  Also, we confirm that the strongest CS emission seen in \cite{Rubio2009} comes from clump no. 2, and there is less intense CS emission in clumps no. 16 and no. 17.
Our clumps no. 11, 12 and 15 have $\frac{I_{^{12}CO}}{I_{^{13}CO}}$ line ratios (10.3, 9.7 and 9.0, respectively) similar to the ratios found by \cite{Rubio2009} for this region, and the rest of the clumps have smaller line ratios, between 3.9 and 7.1.

We resolved sub-parsec size clumps, gathered in groups that coincide with the three ``knots'' reported by \cite{kalari2018} using SEST $^{12}$CO $J=2-1$ observations at $\sim5.6$ pc resolution. Clumps no. 22 and no. 23 coincide spatially and spectrally with KN-1. KN-2 is resolved into three clumps, no. 16, 17 and 18 in our work. Clump no. 16 is the strongest in CO emission of these three clumps and its center velocity coincides with that of the peak emission from KN-2. Clump no. 2 coincides with KN-3 and both are the strongest detections in each of the samples.

The clumps in the diagonal (2, 16, 17, 18, 22 and 23) coincide with the $^{12}$CO dendrogram structures found by \citep{Wong2022-30Dorclumps}. Two clumps have an excellent match: our clump no. 2 matches with their clump no. 57 and clump 23, with their clump 76. Our clumps no. 16, 17 and 18 correspond to their clump no. 67, and the velocity of the $^{12}$CO clump with SCIMES is approximately the average of our CPROPS clumps. 

\subsection{Variations in clump properties with projected distance to R136\label{sec:velgradient}}

\begin{figure}
    \centering
    \includegraphics[width=0.49\textwidth]{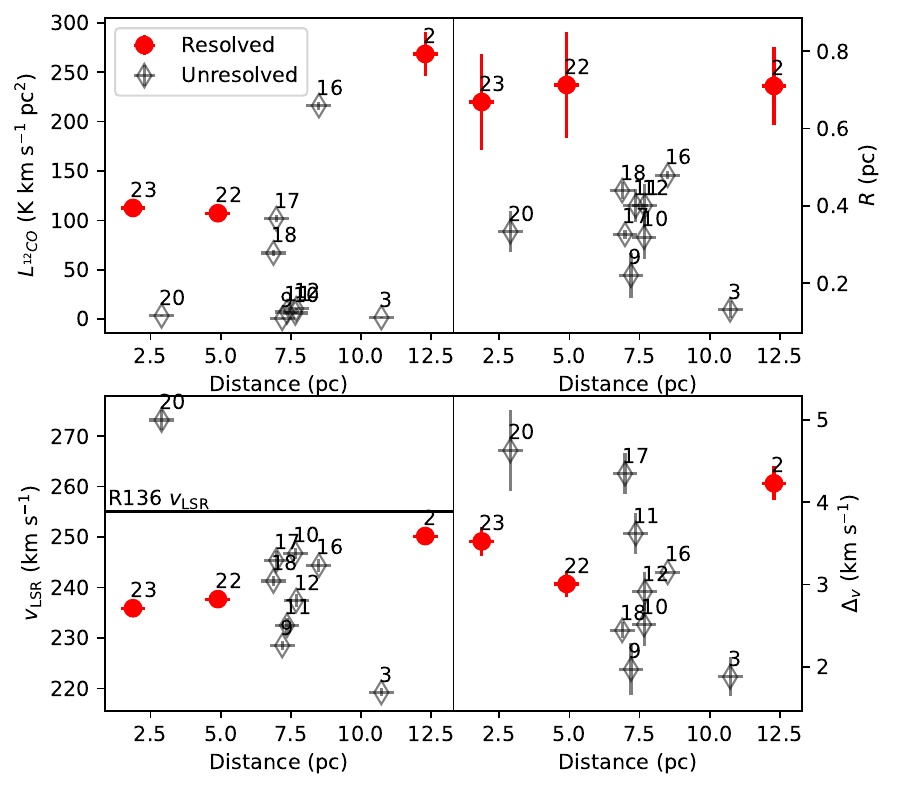}
    \caption{Observed properties of the clumps obtained from $^{12}$CO emission, with respect to projected distance to R136a1. Red points represent the fully resolved clumps from Table \ref{tab:propclouds}, whereas empty black diamonds show the clumps where one axis is unresolved. The black horizontal line represents the \vlsr of the R136 cluster \citep{evans2015}. }
    \label{fig:relwithdistance-cprops}
\end{figure}

We plot the clump $^{12}$CO luminosities, radii, central velocities, and FWHM with respect to the projected distance to the center of the YMC R136 in Fig. \ref{fig:relwithdistance-cprops}. We calculated the projected distance from R136a1, a Wolf-Rayet star which is taken as the center of R136 \citep{Doran2013VLT-Flames}, to the central position of each clump (Table \ref{tab:propclouds}). In general, the clump properties show no clear correlations with distance. However, there is a tendency of increasing \vlsr and increasing luminosity with increasing projected distance for clumps belonging to the diagonal structure seen in Fig. \ref{fig:all-mom0} and Fig. \ref{fig:chanmap-cprops}, which also correspond to the brightest clumps in the sample (2, 16, 17, 18, 22, and 23). 
We note that clump no. 2 has a \vlsr of about 250 km s$^{-1}$, which is close to the R136 cluster velocity (about 255 \kms), as determined from the mean local standard of rest (LSR) radial velocity of the stars in R136 \citep{evans2015}. The rest of the clumps in the diagonal structure are blueshifted with respect to R136. This supports the idea that the clumps in the northwest-southeast diagonal lie slightly in front of R136, and thus between the cluster and us, as suggested by \cite{kalari2018}. The rest of the clumps, which are located outside of the northwest-southeast structure, do not seem to show a tendency in velocity with distance. These clumps are fainter in CO emission that the clumps in the northwest-southeast diagonal structure, and none are resolved.

\begin{figure}
    \centering
    \includegraphics[width=0.49\textwidth]{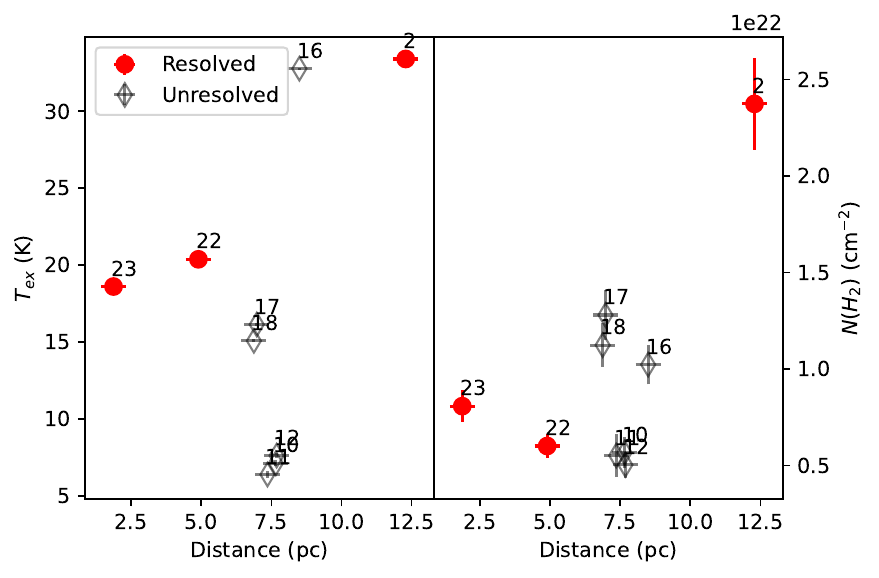}
    \caption{Observed properties of the clumps obtained through LTE analysis, with respect to projected distance to R136a1. The colors represent the same as in Fig. \ref{fig:relwithdistance-cprops}.}
    \label{fig:relwithdistance-lte}
\end{figure}

We plot the excitation temperature and column densities of the clumps obtained with $^{12}$CO and $^{13}$CO in Fig. \ref{fig:relwithdistance-lte}. The resolved clumps (2, 22, and 23) show a tendency of increasing $T_{ex}$ with distance. This tendency is followed by clump no. 16 from the diagonal structure, but not by clumps no. 17 and 18. Clumps to the west of clump no. 18 are $\sim10$ pc away from R136's center, according to the three-dimensional distribution of gas derived in \cite{chevance2016}, whereas the distance of clumps 22 and 23 is uncertain. Therefore, the higher $T_{ex}$  of clump 2 is expected as it is closer to R136 in reality than the other resolved clumps.

In general, our clumps show no strong tendencies with projected distance from R136, except for central velocity. This result is consistent with previous research into 30 Dor clumps, where the $^{12}$CO clump properties do not seem to change with distance to the source of radiation \citep{indebetouw2013, Indebetouw2024H2CO30Dor}. 

\subsection{Scale relations\label{sec:scalerel}}

\begin{figure*}[ht]
    \centering
    \includegraphics[width=0.45\textwidth]{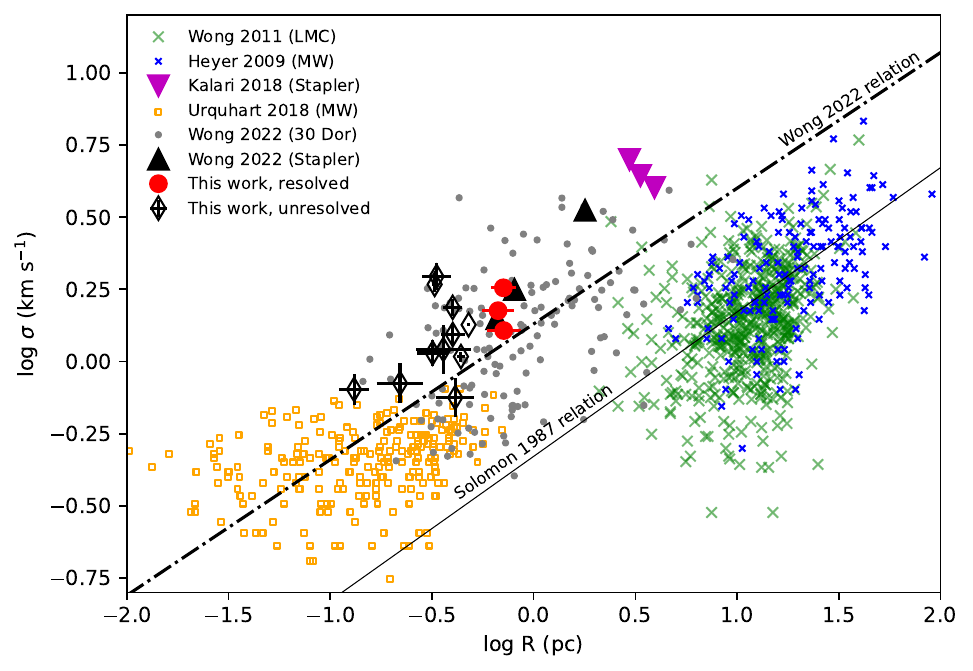}
    \includegraphics[width=0.45\textwidth]{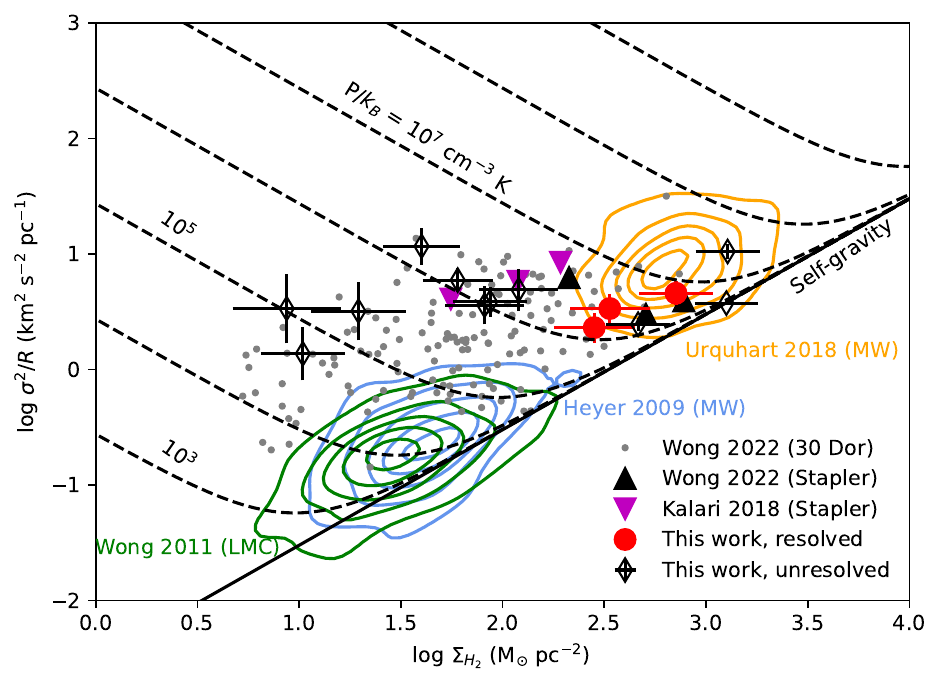}
    \caption{Relationships between physical properties of our sample, together with results in this same region found in \cite{kalari2018}, in 30 Dor from \cite{Wong2022-30Dorclumps} with $\sim1"$ resolution, in the LMC from \cite{wong2011}, in Milky Way clouds from \cite{heyer2009} and inner Galaxy dense clumps from ATLASGAL \citep{Urquhart2018ATLASGAL}. Our CPROPS clumps are shown in red and black in both plots: filled circles represent completely resolved clumps, whereas empty diamonds correspond to clumps that have one unresolved axis. Left: $R$ vs $\sigma_v$ relationship for Molecular Clouds in the LMC and the Milky Way. The black line represents the canonical relation $\sigma_v =0.72R^{0.5}$, followed by the Milky Way clouds \citep{solomon1987}, whereas the dash-dotted line marks the relation found in \cite{Wong2022-30Dorclumps}. Right: $\sigma_v^2/R$ vs $\Sigma_{H_2}$ relationship for Molecular Clouds in different galaxies, including our results and other regions in 30 Dor, the LMC, and the Milky Way. The solid black line represents the approximate $\sigma_v^2/R$ value for increasing $\Sigma_{H_2}$ of a centrally concentrated, virialized clump confined only by self-gravity. Dashed lines represent the relation between $\sigma_v^2/R$ and $\Sigma_{H_2}$ for clumps with $\sigma_v$ affected by self-gravity and external pressure, covering the range $P/k_B=10^3 - 10^9$ cm$^{-3}$ K \citep{Field2011extpress}.}
    \label{fig:relationshipsandKH}
\end{figure*}

Figure \ref{fig:relationshipsandKH} left shows the size-linewidth relation ($\sigma_v-R$) of the $^{12}$CO clumps found via CPROPS, distinguishing between fully resolved clumps and those with unresolved minor axes. Molecular Clouds in virial equilibrium should follow an approximate relation in the form $\sigma_v \propto R^{\alpha}$ \citep{Larson1981}. Clumps in our work, in the LMC, and dense clumps the inner Milky Way populate a different space in the diagram than Galactic Molecular Clouds. We found no significant differences in the $\sigma_v-R$ relation for clumps near R136 compared to other clumps farther from the YMC in 30 Dor, seen by \cite{Wong2022-30Dorclumps} \citep[see also][]{pineda2009, indebetouw2013, nayak2016}. Dense clumps from ATLASGAL follow this relation at smaller scales. We note, however, that these clumps' properties are measured using $^{13}$CO emission, thus we show their values as a reference to the general location they occupy in the $\sigma_v-R$ plot. All structures in 30 Dor, including our sample, follow a power law similar to Milky Way clouds, but with higher velocity dispersions relative to both the Milky Way \citep{heyer2009} and other LMC clouds excluding 30 Dor \citep{wong2011}.

The offset of the $\sigma_v-R$ relation is dependent on the mass surface density $\Sigma$ of the clumps and thus indicates how bound they are \citep{heyer2009}. If clumps are virialized and confined by self-gravity, a $\sigma_v^2/R - \Sigma$ plot should follow a power law with a scaling of $ \frac{\pi G}{5}$. \citep{kalari2018} suggests that the larger linewidth of the Stapler nebula clumps are a reflection of the external pressure necessary to bound them \citep{Field2011extpress}. In Fig. \ref{fig:relationshipsandKH} (right), we plot $\sigma_v^2/R - \Sigma$ using $\Sigma_{H_2}$ obtained from $^{12}$CO emission (Table \ref{tab:propclouds}). The resolved clumps are closer to the self-gravity line than those identified by \cite{kalari2018}, but still lie above it within uncertainties. They remain consistent with external pressures of $10^6-10^7$ cm$^{-3}$ K \citep[based on][]{Field2011extpress}, aligning with the gas pressure of $\sim(0.85-1.2)\times10^{6}$ cm$^{-3}$ K found in the region \citep{chevance2016}. Unresolved clumps position above the self-gravity line, consistent with clumps found by \cite{Wong2022-30Dorclumps}. All 30 Dor clumps, together with dense clumps from ATLASGAL, lie farther from the self-gravity line than LMC clouds \citep{wong2011} and the Milky Way \citep{heyer2009}. In general, our clumps appear either unbound or bound by a combination of gravity and external pressure, similar to other clumps in 30 Dor \citep{indebetouw2013}.

The origin of the elevated $\sigma_v$ and external pressure for 30 Dor clumps can be attributed to radiation pressure and/or gas cloud collisions. Recent studies have shown that the formation of R136 was triggered by the fast ($\sim100$ \kms) collision of two HI flows \citep{Fukui2017HIcol, Maeda2021collisionsmodel}.
Our clumps are located in what corresponds to the bridge gas layer formed by this collision \citep[the I-component in][]{Tsuge2024IcompHIflow}, and the velocity range of all our clumps ($\sim50$ \kms difference) is consistent with both the velocity difference between the colliding clouds in \cite{Fukui2017HIcol}, and the velocity range of the bridge. This suggests that the higher $\sigma_v$ of the 30 Dor clumps originates from this initial collision, which has been suggested to come from the tidal interaction of the LMC with the SMC \citep{Fujimoto1990LMCSMCinteraction}. Nevertheless, the external pressure onto the gas by the radiation of massive stars is an important factor for the compression of these clumps, given their proximity to R136 \citep{chevance2016}. Although the external gas pressure by collision found in \cite{Tsuge2024IcompHIflow} is in the order of $10^6$ K cm$^{-3}$, similar to the gas pressure found by \cite{chevance2016}, \cite{Lopez2011radpressureR136} found that the radiation pressure is also around $10^6$ K cm$^{-3}$ and dominates the gas pressure within a few tens of pc from R136. It is possible that both compression and radiation pressure shape the equilibrium state of these clumps.

In summary, the Stapler nebula clumps exhibit properties similar to other clumps in 30 Dor and massive dense clumps in the Milky Way, showing increased turbulence compared to typical LMC clouds but comparable to clumps elsewhere in 30 Dor. This suggests that dense clumps in 30 Dor are not substantially different from clumps in other regions except for increased levels of turbulence, consistent with findings by \cite{indebetouw2013, Indebetouw2020dendog30Dor} and \cite{Grishunin202430DorLMCcomp}. Despite theoretical models predicting molecular gas evacuation within $10-15$ pc of R136 \citep{dale2012}, clumps approximately 10 pc from R136's center \citep{chevance2016} not only survive but have similar turbulence levels as clumps farther away and form stars (Sect. \ref{sec:ysocomparison}), suggesting that YMC feedback may not be so effective in disrupting surrounding gas and subsequent star formation \citep[see][and references within]{Krumholz2019rev}.

\subsection{Comparison between clumps in R136 and regions in the LMC and Milky Way}\label{sec:lineratiocomparison}

We investigated if the physical and chemical properties of our clumps are comparable to those in other regions of the LMC and of massive dense clumps in our Galaxy.
We compared the properties and line ratios of clumps near R136 with other clumps in the LMC, including 30 Dor \citep{indebetouw2013,minamidani2011,Rubio2009}, N113 \citep{seale2012,paron2014}, N159 \citep{minamidani2011,paron2016}, and N11 \citep{celispena2019}. N113 and N159 are the only other regions where $^{13}$CO, CS, HCO$^+$, and HCN have been detected in the 345 GHZ window in the LMC. We compared also with dense clumps found in the inner Galaxy from ATLASGAL \citep{Urquhart2018ATLASGAL} and CHIMPS \citep{Rigby2019CHIMPS} surveys. We note that some of these works detect CO in a lower excitation transition \citep[e.g., $^{12}$CO$3-2$ instead of 2-1,][]{indebetouw2013}, which require, for instance, higher excitation temperatures. 

The excitation temperatures ($T_{ex}$) obtained from $^{12}$CO $J=3-2$ emission in our sample are lower than $T_{ex}$ sampled in other 30 Dor regions but comparable to less bright \ion{H}{II} regions in the LMC. The range of $T_{ex}$ is consistent, albeit reaching slightly higher temperatures in our resolved clumps (about 10 K more) than the $T_{ex}$ distribution in clumps within the inner Galactic plane \citep{Rigby2019CHIMPS}. Clump no. 2 shows the highest $T_{ex}$ at 33.39 K, while 30Dor-10 regions exhibit $T_{ex}$ between $40-60$ K \citep{indebetouw2013}. This difference is expected, as $^{12}$CO $J=3-2$ emission in 30Dor-10 exceeds $^{12}$CO $J=2-1$ emission near R136 by a factor of five \citep{kalari2018}. Clumps no. 17, 18, 22, and 23 have $T_{ex}$ values similar to N113 \citep[$\sim20$ K][]{paron2014}, while clumps no. 8, 10 to 12, 14, and 15 show values (7-13 K) comparable to N11 clumps \citep{celispena2019}. The lower temperatures are likely a result of a low beam filling factor, due to the unresolved nature of our clumps.

Column densities $N(H_2)$ and $N(^{13}CO)$ in our clumps are lower than in other 30 Dor clumps, though the brightest clump in our sample have densities comparable to other LMC regions. $N(^{13}CO)$ in 30Dor-10 reaches $(1-5)\times10^{16}$ cm$^{-2}$ \citep{indebetouw2013}, an order of magnitude higher than our study. Only clump no. 2, the farthest from R136 in projection and one of few resolved clumps, has comparable densities with $N(^{13}CO)=1.32\pm0.13\times10^{16}$ cm$^{-2}$ and $N(H_2)= 2.23\times10^{22}$ cm$^{-2}$. The northwest-southeast diagonal structure clumps (no. 2, 16, 17, 18, 22, and 23) have $N(H_2)$ between $(0.6 - 2.4)\times10^{22}$ cm$^{-2}$, similar to $N(H_2)$ peak column densities found in the N11 region \citep{celispena2019} and within the range observed by \cite{Urquhart2018ATLASGAL} in Milky Way clumps. Lower column densities do not necessarily indicate lower volume densities. Based on LVG modeling and $\frac{I_{^{12}CO}}{I_{^{13}CO}}$ line ratios, our clumps have estimated volume densities $n(H_2)\approx10^3-10^4$ cm$^{-3}$. The detection of CS and HCN, which require densities of at least $\sim 10^7$ cm$^{-3}$, suggests density increases inside the clumps that our observations cannot fully resolve.

Our $\frac{I_{^{12}CO}}{I_{^{13}CO}}$ line ratios ($3.9-10.3$) align with those found in other 30Dor clumps \citep[$5.5-8.2$ but with $\sim10$ pc resolution,][]{minamidani2011} and N11 \citep[$6.5-13$,][]{celispena2019}. Only clumps no. 11 and 12 show higher ratios (10.3 and 9.7), while clumps no. 2 and 8 have ratios below 5. These ratios are also consistent with N159 and N113 \citep[$5.5-8.2$][]{minamidani2011, paron2016, paron2014}, with clumps no. 11 and 12 showing ratios similar to less bright \ion{H}{II} regions N132 and N166 \citep[10.31 and 11.14, respectively]{paron2016}.

The $\frac{I_{CS}}{I_{^{13}CO}}$ ratios in most of the northwest-southeast diagonal structure are similar to N113 \citep[$\sim0.06$,][]{paron2014}, except clumps no. 22 and 23, which have lower ratios (0.02 and 0.05) and are most similar to N159 \citep[$\sim0.03$,][]{paron2016}. Assuming both molecules' emission are optically thin, the lower ratio potentially indicates lower densities. Clump no. 10 has a notably higher ratio (0.22), suggesting higher density. The $\frac{I_{HCN}}{I_{^{13}CO}}$ ratios indicate a similar tendency: $\frac{I_{HCN}}{I_{^{13}CO}}$ ratios in most of our clumps ($0.11-0.34$) are $2-6$ times higher than in N159 and N113 \citep[$0.06-0.07$][]{paron2014, paron2016}, while clumps no. 22 and 23 show lower ratios (0.02 and 0.04), consistent with their lower derived column densities $N(H_2)$ (Table \ref{tab:textauandN}).
The $\frac{I_{HCO^{+}}}{I_{HCN}}$ ratios (mostly between $3.5-6.3$) are comparable to N113 (about 4.8) but lower than N159 \cite[about 8.7,][]{paron2016}, with clump no. 22 being an exception (12.2). Lower $\frac{I_{HCO^{+})}}{I_{HCN}}$ ratios correlate with higher volume density and star formation, as seen in clumps no. 10 and 17 (ratios of 3.51 and 4.62) which are associated with young stellar objects (YSOs). This is consistent with \cite{seale2012}, where a lower HCO$^{+}$/HCN ratio (both in the $J=1-0$ transition) correlates to a higher volume density.

In summary, molecular clumps near R136 share similar chemical and physical properties with other LMC regions. We suggest the Stapler nebula is eroded by photoionizing radiation from R136, with only the densest clumps surviving, as density can hinder the dispersal of gas \citep{dale2012, Krumholz2019rev}. Once these clumps withstand photoionization, protected by their density and possibly bound by external pressure, they develop similarly to other Molecular Cloud clumps.

\subsection{Gas and dust emission with respect to the LMC\label{sec:gasmassdiscussion}}

Our analysis in Sect. \ref{sec:coandcontcomparison} revealed that areas covered by $^{12}$CO emission exceed those covered by the same clumps in 0.88 mm continuum. This contradicts expectations for low metallicity environments, where a larger H$_2$ envelope not traced by $^{12}$CO should be better traced by dust \citep{bolatto2013}. This differs from findings in the N11 region, where dust and $^{12}$CO emissions have similar extensions \cite{herrera2013}, and from the SMC, where $^{12}$CO emission covers smaller areas than 1.2 mm continuum \citep{rubio2004,bot2007}. These discrepancies likely stem from observational limitations. Both continuum and $^{12}$CO images show negative emission bowls, particularly visible in $^{12}$CO channel maps (Fig. \ref{fig:chanmap-cprops}), indicating missing flux at extended scales. Additionally, the 0.88 mm ALMA continuum has significant free-free emission contribution (Sect. 4.4.1). Total Power (TP) observations would be necessary to detect extended continuum emission.

\begin{figure}
    \centering
    \includegraphics[width=0.95\linewidth]{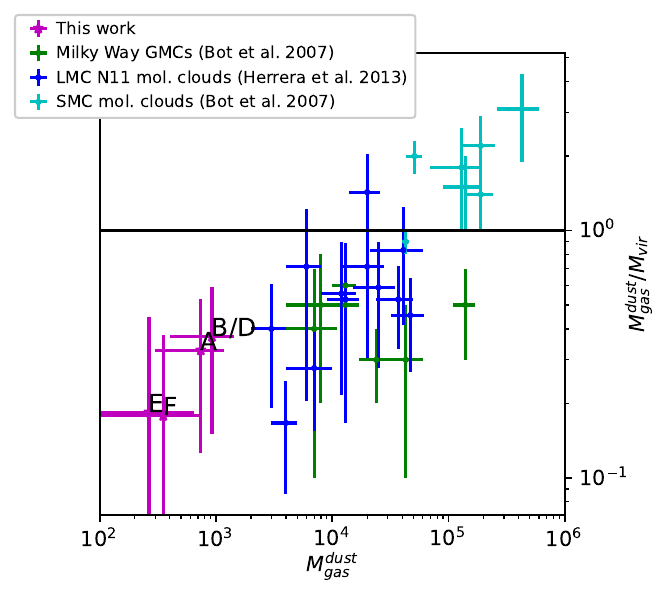}
    \caption{$M_{gas}^{dust}$ plotted against the ratio $M_{gas}^{dust}/M_{vir}$. We include the ratios found by \cite{bot2007} in the SMC in cyan, \cite{herrera2013} in N11 in blue, and Milky Way values calculated in \cite{bot2007} in green. Clumps are labeled according to their 0.88 mm continuum labels. The horizontal black line marks the $M_{gas}^{dust}/M_{vir}=1$ equality.}
    \label{fig:mgasmdustmvirratio}
\end{figure}

The mass ratios $M_{vir}/M_{gas}^{CO}$, $M_{vir}^{A_{880\mu m}}/M_{gas}^{dust}$, and $M_{gas}^{CO,A_{880\mu m}}/M_{gas}^{dust}$ are comparable to those found in the LMC but differ from SMC results. Our ratios align with those from N11, the LMC's second brightest nebula \citep{herrera2013}. The observation that $M_{gas}^{CO,A_{880\mu m}}/M_{gas}^{dust}\lesssim1$ (except for clump A) while $M_{vir}/M_{gas}^{CO,A_{880\mu m}}>1$ for resolved CO clumps supports the possibility that we're not sensitive to the complete dust emission. Nevertheless, given the large uncertainties in $M_{gas}^{dust}$,, the differences between $M_{vir}$, $M^{CO}_{\mathrm{gas}}$ and $M_{gas}^{dust}$ values fall within a factor of $2-3$, indicating rough consistency (Table \ref{tab:Mcocont}). These variations may be explained by our assumptions for $\alpha_{CO}$ in calculating $M^{CO}_{\mathrm{gas}}$, for $\kappa$ in determining $M_{gas}^{dust}$, and the virial equilibrium assumption for $M_{vir}$.

Figure \ref{fig:mgasmdustmvirratio} compares our $M_{vir}^{A_{880\mu m}}/M_{gas}^{dust}$ values with those from the LMC \citep{herrera2013}, SMC, and Milky Way \citep{bot2007}. Our results differ from the SMC, where virial masses are lower than dust-derived gas masses, and align with the Milky Way and LMC patterns, where virial masses are larger. All LMC clumps in the sample fall below the $M_{gas}^{dust}/M_{vir}=1$ line. This may relate to the SMC's lower metallicity \citep[about 1/5\,$Z_{\odot}$,][]{lee2005} compared to the LMC 1/5\,$Z_{\odot}$), causing CO areas to shrink and trace less mass \citep[][and references within]{bolatto2013}. A limitation of this interpretation is the difference in beam sizes between our study (4.7\arcsec) and \cite{bot2007} (24\arcsec), so we are not necessarily comparing the same kind of structures (clouds instead of clumps).

Reducing uncertainty in $M_{gas}^{dust}$ would improve our $M^{CO}_{\mathrm{gas}}/M_{gas}^{dust}$ and $M_{vir}/M_{gas}^{dust}$ ratio estimates. This requires multiple images across mm to cm wavelengths to properly fit dust and free-free emission, along with deeper ALMA observations and single-dish data to recover extended emission.

\subsection{Association of gas and dust with YSOs \label{sec:ysocomparison}}

\begin{figure}[ht]
    \centering
    \includegraphics[width=0.5\textwidth]{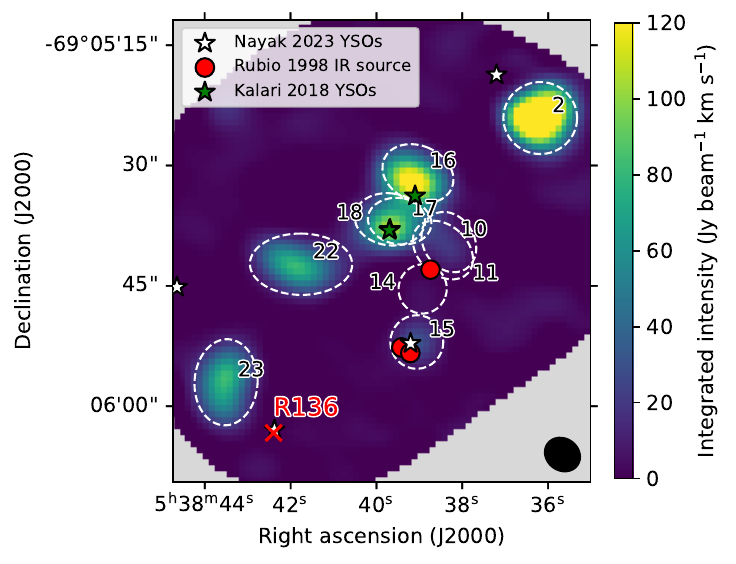}
    \caption{Velocity integrated image of $^{12}$CO emission between 235 and 250 km s$^{-1}$. White dashed ellipses represent the position and sizes of the major and minor axes of clumps found in Sect. \ref{sec:cprops}. The black labels over the ellipses correspond to the ID number of each clump. White stars indicate the YSO from \cite{Nayak2023YSOtarantula} catalog. Green stars indicate the YSOs identified in \cite{kalari2018}. Red points indicate the IR sources from the \cite{rubio1998} catalog. The black ellipse in the bottom left corner represents the beam size.}
    \label{fig:ysoversusclouds}
\end{figure}

The Stapler nebula shows active star formation, in particular in the middle of the diagonal structure \citep{kalari2018}. We determined which of our clumps are correlated with YSOs and embedded infrared sources. Figure \ref{fig:ysoversusclouds} shows the velocity integrated image between 235 and 250 km s$^{-1}$ of the $^{12}$CO line emission cube, together with the clumps obtained using CPROPS that lay in the diagonal structure plus clumps 10, 11, 14 and 15, recognized YSO candidates in the region from \cite{Nayak2023YSOtarantula} and \cite{kalari2018}, and infrared sources from \cite{rubio1998}. There are several identified YSO in our field of view, but only three YSO candidates and one IR source coincide with our observed $^{12}$CO emission and are all concentrated in the central region of our observations. The three YSO candidates coincide with clumps detected in HCN, which has the highest critical density of our sample (Table \ref{tab:cubeprops}), and thus associated with some of the densest clumps in the region.

The YSO candidate Y26 from \cite{Nayak2023YSOtarantula} is associated with clump no. 15/continuum source C. This YSO was first identified as a YSO candidate in \cite{gruendl2009} (source 053839.24-690552.3), and coincides with two NIR sources, IRSW-118 and IRSW-125 \citep{rubio1998}, but they could not associate the candidate with one specific NIR source due to the resolution of Spitzer's IRAC images (2\arcsec). Clump no. 15 has the largest contribution of free-free emission to the 880 continuum ALMA flux as presented in Sect. \ref{sec:freefree}: 
the source has been associated with an optical counterpart, P702 \citep{parker1993}, classified as an O9.5 type star. Therefore, this YSO could be the ionizing source that produces the higher free-free emission associated to this clump with respect to the rest of the region \citep{walborn2014}.

\cite{kalari2018} identified two YSO candidates (KN2-A and KN2-B) in the region they labeled Knot 2, which we resolve into three clumps (16 to 18). The properties of the clumps are consistent with the difference in developments stage of these two YSO, evaluated using the \cite{Robitaille2017SED} spectral energy distribution (SED) fitting and the SED slope in the IR. Clump 16 coincides with KN2-B, classified as a massive Class 0 protostar by \cite{kalari2018}. This clump is the brightest in $^{12}$CO and $^{13}$CO emission of the three clumps, and also shows CS and HCN emission, which correlates with this protostar being the most embedded. On the other hand, KN2-A \citep[corresponding to IR source IRSW-127 in][]{rubio1998}, which is associated with clumps 17 and 18 in our work, is classified as Class II, older than KN2-B and not embedded. Thus, this YSO is possibly outside these clumps, corroborating the picture of Knot 2 from \cite{kalari2018}. 

There is an IR source in the intersection of clumps 10, 11 and 14. This source, IRSW-105, has very little IR excess in the $(J-H)$ v/s $(H-K_s)$ color-color diagram and its colors are consistent with a highly reddened O3 V type star \citep{rubio1998}. It is possible that the clumps are in front of these source and are responsible for its reddening.

There are other YSO in the field of view, but these are not associated with clumps in this work. Clump 2 was classified as a starless dust core in \cite{gruendl2009}. There is a YSO candidate from \cite{Nayak2023YSOtarantula} approximately 2 pc away from the clump center, but it does not fall within the clump radius.


\section{Summary and conclusions\label{sec:summary}}

In this work, we present ALMA Band 7 observations with $\sim1$ pc resolution. Our ALMA map detects 24 $^{12}$CO $J=3-2$ clumps, the brightest and densest of which lie in the region identified as the ``Stapler Nebula'' by \cite{kalari2018}, seen as a dark lane in previous HST observations. New CO clumps are identified outside of this region, one of which shows star formation activity (clump no. 15). The molecular clumps are found at a projected distance between 2 and 13 pc from R136 and several these clumps are barely resolved at our resolution. We detect for the first time HCO$^{+}$ $J=4-3$ and HCN $J=4-3$ in these clumps. We summarize the main results of this work in the following.

Detection of all the molecular species observed , $^{13}$CO $J=3-2$, CS $J=7-6$, HCO$^{+}$ $J=4-3$, and HCN $J=4-3$  was found in only six clumps of the 24 $^{12}$CO $J=3-2$ clumps. These six clumps are located in the northwest-southeast diagonal optical dark lane. CS $J=7-6$ and HCN $J=4-3$ emission in these clumps suggests high density ($10^{6}$ cm$^{-3}< n < 10^{8}$ cm$^{-3}$). Using our results for $^{12}$CO $J=3-2$ and $^{13}$CO $J=3-2$ molecular emission, we calculated the excitation temperatures $T_{ex}$ and the H$_2$ column densities $N(H_2)$ of our clumps, assuming LTE. The resulting $N(H_2)$ for our sample ranges from 4.2 to $23.7\times10^{21}$ cm$^{-2}$. 
The dust continuum map at 0.88 mm shows emission with a similar spatial distribution as the molecular line emission. 
The continuum emission mapped by ALMA is dominated by the free-free emission, contributing between 50 and 85\% of the total 0.88 mm continuum flux, as expected from an \ion{H}{II} region.

We compared the clump masses obtained through the different methods. The virial masses, gas masses derived from CO luminosity and also from dust emission were determined in the same area for each cloud observed in continuum emission to make the comparison. We found that the virial masses in this case are approximately two to five times larger than the gas masses obtained from dust emission in the same area, which is consistent with previous findings in the LMC. Virial masses are the highest because the velocity dispersion of the clumps is a measure of both the mass and external gas pressure. The $M_{vir}/M_{gas}^{dust}$ ratios are similar to these comparisons in other regions in the LMC. Gas masses obtained from dust emission are similar to the gas masses obtained from CO luminosity, when using a fixed conversion factor of $\alpha_{^{12}CO}=8.4\pm3.0$ M$_{\odot}$ (K km s$^{-1}$ pc$^{2}$)$^{-1}$. 

The line ratios of the clumps are similar to line emission ratios found in studies in 30 Dor and other \ion{H}{II} regions in the LMC. This suggests that the physical and chemical properties 
of the studied clumps near R136 are similar to other molecular clumps located in 30 Dor farther away from R136. We suggest that these clumps are the densest parts of a Molecular Cloud that has been carved by the strong radiation from R136, reflected in the fact that they are detected in molecular transitions with high critical densities. These high densities allow for the birth of a new generation of stars, at distances less than 20 pc away from R136. These results present constraints on the physical conditions in which dense star-forming molecular gas is affected (or not) by ionizing and mechanical feedback of early-type stars, which can further inform newer models of stellar feedback affecting star formation.

Deeper ALMA observations, together with Total Power array observations, would allow to resolve the molecular clumps in 0.88 mm continuum. Further work using non-LTE models such as RADEX could be applied to better quantify the densities volume densities and kinetic temperatures  of the clumps found in this work.


\begin{acknowledgements}

We thank the anonymous referee for their constructive comments on the manuscript. We thank Alejandra Meza for helping with the CO clump cross-matching to \cite{Wong2022-30Dorclumps} clumps. M.T.V-M. gratefully acknowledges financial support from ANID (CONICYT) through the scholarship CONICYT-PFCHA/Magister Nacional 2018 – 22180279, support from the Max Planck Society, and sponsorship provided by the European Southern Observatory through the Research Fellowship program. M.R. wishes to acknowledge support from ANID(CHILE) through FONDECYT grant No 1190684 and partial support from Basal FB210003.  This paper makes use of the following ALMA data: ADS/JAO.ALMA\#2017.1.00368.S. ALMA is a partnership of ESO (representing its member states), NSF (USA) and NINS (Japan), together with NRC (Canada), MOST and ASIAA (Taiwan), and KASI (Republic of Korea), in cooperation with the Republic of Chile. The Joint ALMA Observatory is operated by ESO, AUI/NRAO and NAOJ. The National Radio Astronomy Observatory is a facility of the National Science Foundation operated under cooperative agreement by Associated Universities, Inc. This research made use of Astropy (http://www.astropy.org), a community developed
core Python package for Astronomy.

\end{acknowledgements}

%
%
\bibliographystyle{aa}
\bibliography{aa54327-25}

\begin{appendix}

\onecolumn

\section{Gaussian fitting method\label{sec:cpropssecondstep}}

We obtained the properties of clumps labeled 7, 8, 16, 17, and 18 in the final catalog (Table \ref{tab:cpropsresults}) using Gaussian fitting. Clumps 7 and 8 were identified by CPROPS as one clump, whereas clumps 16, 17 and 18, although identified as separate sources, are blended along the line of sight.

\begin{figure*}[htbp]
	\begin{center}
		\includegraphics[width=0.7\textwidth]{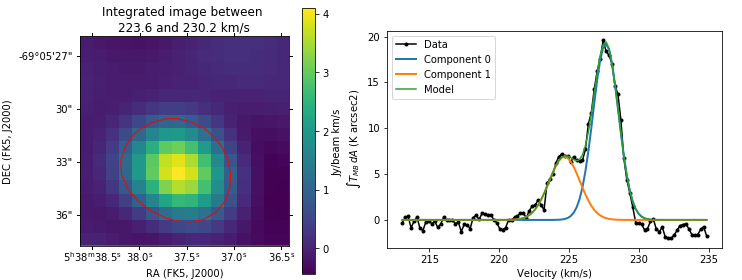}
		\includegraphics[width=0.7\textwidth]{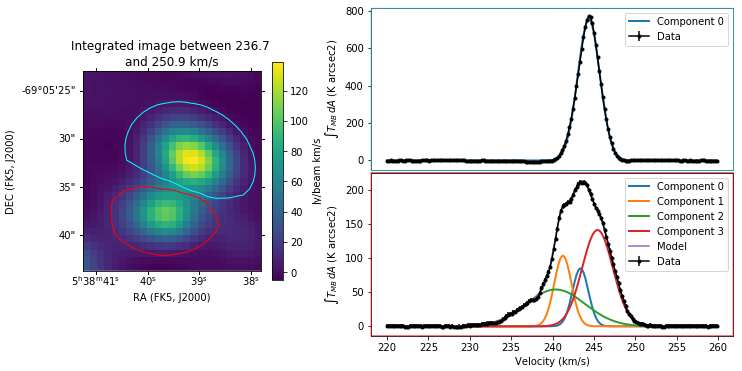}
		
		\caption{Velocity integrated $^{12}$CO images in the locations of the 5 clumps we characterize manually (no. 7, 8 , 16, 17 and 18), together with the integrated spectra in the delimited areas. Black points in the spectra correspond to the line emission data and colored lines correspond to the Gaussian fits to each clump. Top: $^{12}$CO emission integrated between 223.6 and 230.9 km/s (left), together with the integrated spectra in the red region in the integrated image (right). Gaussian component 0 (blue line) corresponds to emission from clump no. 8, and component 1 (orange line) to clump no. 7. Bottom: $^{12}$CO emission integrated between 236.7 and 250.9 km/s (left), together with the integrated spectra in the cyan and red regions marked in the integrated image (right). The integrated spectra at the top right corresponds to the spectra inside the cyan area. The integrated spectra at the bottom right corresponds to the one inside the red area. Gaussian component 0 (blue lines) from both spectra show line emission from clump no. 16 from Table \ref{tab:cpropsresults}. Gaussian component 2 (green line) from the bottom right spectra corresponds to emission from clump no. 22. Components 1 and 3 (orange and red lines) from the bottom right spectra correspond to emission from clumps no. 18 and no. 17, respectively. \label{fig:clouddistinc}}
	\end{center}
\end{figure*}

For clumps 7 and 8, we first integrated the $^{12}$CO line cube between 223.6 and 230.2 km s$^{-1}$. We then obtained the integrated spectrum by adding up all spectra inside the image's 3$\sigma$ contour (for this integrated image, $\sigma = 0.3$ \Jyb \kms). The integrated spectrum (Fig. \ref{fig:clouddistinc} top right) is best fit with two Gaussian components. We thus consider this source as two different clumps which are separated in velocity but are located in the same line of sight. Component 0 corresponds to clump no. 8 and component 1, to clump no. 7 from Table \ref{tab:cpropsresults}. 
To determine the properties of clumps 7 and 8, we integrated the $^{12}$CO line cube in the range $(v-2\sigma_v, v+2\sigma_v)$ for each clump, where $v$ and $\sigma_v$ are the central velocity and dispersion obtained for the individual Gaussian components (reported in Table  \ref{tab:cpropsresults})). From these images we determined the central position and the major and minor axes of the clumps using the same moment method employed in CPROPS, described in \cite{rosolowsky2006}. The resulting sizes and luminosities are reported in Table \ref{tab:cpropsresults}.

For clumps 16, 17, and 18, first we integrated the $^{12}$CO line cube in the velocity range between 236.7 and 250.9 km/s, where we identified two sources, as shown in the bottom left panel of Fig. \ref{fig:clouddistinc}. We obtained the spectra of each source by adding up all the spectra inside the area of each clump, defined as the 3$\sigma$ contour of the integrated intensity ($\sigma = 2.6$ \Jyb \kms). The spectra of the northwestern source (encircled in cyan), shown in Fig. \ref{fig:clouddistinc} center right, is well fitted using one Gaussian component with a central velocity of  $v=244.40$ \kms and a velocity dispersion $\sigma_v=1.34$ \kms. This source is labeled as clump 16. On the other hand, the spectra of the southern source (encircled in red), shown in Fig. \ref{fig:clouddistinc} bottom right, is best fitted with 4 Gaussian components. 
Gaussian components 1 and 3 (the orange and red lines in Fig. \ref{fig:clouddistinc} bottom right) correspond to clumps 18 and 17, respectively, identified by CPROPS. Components 0 and 2 correspond to clumps 16 and 12, respectively, identified through visual inspection of the $^{12}$CO line cube. We determined the properties of clumps 16, 17, and 18 using the same procedure as for clumps 7 and 8.

\section{CPROPS detections in $^{13}$CO, CS, HCO$^+$, and HCN \label{appendix:cpropsmolecules}}

Tables \ref{tab:13co32cprops} to \ref{tab:hcn43cprops} list the detected clumps in the $^{13}$CO, CS, HCO$^+$, and HCN ALMA line cubes, with the same parameters as those stated in Sect. \ref{sec:cprops}. We selected those emissions that, according to CPROPS, have $S/N>5$, and manually inspected the results to leave out false detections. We then matched the emissions found in each molecule with its corresponding clump detected in the $^{12}$CO line cube, according to their position ($\alpha,\delta$) and central velocities $v$. We report the properties obtained through CPROPS for each molecule: $\alpha$, $\delta$, $v$, velocity FWHM $\Delta v$, peak temperatures $T_{peak}$, deconvolved radii $R_{dc}$ and not deconvolved radii $R$, and luminosities of each molecular species.

The radii were calculated from the second moments of emission along the major and minor axes of the clumps, delivered by CPROPS. For unresolved clumps (i.e., one or both of the second moments of emission of the clump are smaller than $\sigma_{beam}$, Sect. \ref{sec:physprops}), we used Equation \ref{eq:nonresolvedradii} to calculate the upper limit to the deconvolved radius.

In the case of clump no. 11, we added a detection in $^{13}$CO which is not detected by CPROPS but is detected after a manual inspection of the $^{13}$CO line cube. We do not attempt to characterize its size in this work.

\begin{table*}[!h]
\caption{Molecular clumps detected in the $^{13}$CO line cube.\label{tab:13co32cprops}}

\centering
\begin{tabular}{lllllllll}
\hline\hline
ID &  $\alpha$  &    $\delta$ &  $T_{peak}$ &     $v$ &   $\Delta v$ &    $R_{dc}$ &    $R$ & $L_{^{13}CO}$ \\
 & (J2000) & (J2000) & (K) & (km s$^{-1}$) & (km s$^{-1}$) & (pc) & (pc) & (K km s$^{-1}$ pc$^2$)  \\
\hline
2      &  05:38:36.2 &  -69:05:24.0 &   8.13 &  $250.1$ &  $3.5\pm0.2$ &   $0.42\pm0.06$ &  $1.01\pm0.13$ &  $56.14\pm5.08$ \\
8$^*$         &  05:38:37.6 &  -69:05:33.7 &   0.28 &  $227.8$ &  $1.5\pm0.3$ &   $0.16\pm0.04$ &  $0.74\pm0.23$ &   $0.56\pm0.05$ \\
9$^{\dagger}$         & 05:38:37.9 &  -69:05:42.6 &   0.19 &  $228.3 $ &  - &     - &   - &     - \\
10$^*$         &  05:38:38.2 &  -69:05:40.0 &   0.47 &  $246.6$ &  $2.0\pm0.4$ &   $0.23\pm0.05$ &  $0.78\pm0.21$ &   $1.12\pm0.11$ \\
11$^{\dagger}$        &  05:38:38.6 &  -69:05:41.1 &   0.15 &  $232.3$ &  - &     - &   - &     - \\
12$^*$         &  05:38:38.6 &  -69:05:38.7 &   0.31 &  $235.5$ &  $7.9\pm0.6$ &   $0.50\pm0.09$ &  $0.94\pm0.17$ &   $1.78\pm0.11$ \\
14$^*$         &  05:38:39.0 &  -69:05:45.6 &   0.42 &  $241.5$ &  $1.5\pm0.2$ &   $0.09\pm0.02$ &  $0.76\pm0.22$ &   $0.86\pm0.09$ \\
15$^{**}$         &  05:38:39.1 &  -69:05:51.8 &   0.70 &  $247.8$ &  $2.4\pm0.3$ &  $<0.85\pm0.15$ &  $0.81\pm0.18$ &   $2.52\pm0.22$ \\
16        &  05:38:39.2 &  -69:05:32.0 &   5.20 &  $244.3$ &  $2.2\pm0.2$ &   $0.42\pm0.07$ &  $0.99\pm0.16$ &  $21.73\pm2.31$ \\
17        &  05:38:39.6 &  -69:05:37.2 &   2.05 &  $246.3$ &  $2.5\pm0.2$ &   $0.54\pm0.09$ &  $1.02\pm0.17$ &   $7.87\pm1.00$ \\
18        &  05:38:40.0 &  -69:05:37.4 &   1.15 &  $240.9$ &  $2.0\pm0.2$ &   $0.54\pm0.09$ &  $1.03\pm0.17$ &   $4.75\pm0.53$ \\
21$^{**}$        &  05:38:41.5 &  -69:05:25.3 &   0.14 &  $254.5$ &  $1.0\pm0.4$ &  $<0.85\pm0.34$ &  $0.62\pm0.40$ &   $0.12\pm0.09$ \\
22$^{*}$        &  05:38:41.9 &  -69:05:42.4 &   2.41 &  $237.8$ &  $2.4\pm0.2$ &   $0.72\pm0.10$ &  $1.11\pm0.14$ &  $14.52\pm1.40$ \\
23        &  05:38:43.5 &  -69:05:57.3 &   2.90 &  $236.0$ &  $2.6\pm0.2$ &   $0.55\pm0.08$ &  $1.05\pm0.14$ &  $17.37\pm1.70$ \\
\hline

\end{tabular}
\tablefoot{
\tablefoottext{*}{Minor axis is unresolved.}
\tablefoottext{**}{Both axes are unresolved.}
}
\end{table*}

\begin{table*}[!h]
\caption{Molecular clumps detected in the CS line cube\label{tab:cs76cprops}}
\centering
\begin{tabular}{lllllllll}
\hline\hline
ID &  $\alpha$  &    $\delta$ &  $T_{peak}$ &     $v$ &   $\Delta v$ &    $R_{dc}$ &    $R$ & $L_{CS}$ \\
 & (J2000) & (J2000) & (K) & (km s$^{-1}$) & (km s$^{-1}$) & (pc) & (pc) & (K km s$^{-1}$ pc$^2$)  \\
\hline
2         &  05:38:36.5 &  -69:05:24.3 &   0.64 &  $249.0$ &  $2.0\pm0.2$ &   $0.17\pm0.03$ &  $0.86\pm0.18$ &  $2.17\pm0.22$ \\
10$^{**}$        &  05:38:38.3 &  -69:05:40.2 &   0.14 &  $246.5$ &  $1.3\pm0.3$ &  $<0.85\pm0.21$ &  $0.76\pm0.25$ &  $0.23\pm0.03$ \\
16$^{*}$        &  05:38:39.2 &  -69:05:32.5 &   0.45 &  $244.0$ &  $1.8\pm0.2$ &   $0.29\pm0.05$ &  $0.82\pm0.18$ &  $1.20\pm0.13$ \\
17$^{*}$       &  05:38:39.7 &  -69:05:37.8 &   0.20 &  $245.4$ &  $2.7\pm0.3$ &   $0.48\pm0.09$ &  $0.92\pm0.19$ &  $0.81\pm0.05$ \\
18$^{*}$        &  05:38:40.0 &  -69:05:38.6 &   0.09 &  $241.1$ &  $1.4\pm0.3$ &   $0.14\pm0.04$ &  $0.83\pm0.26$ &  $0.16\pm0.03$ \\
22$^{*}$        &  05:38:42.0 &  -69:05:41.5 &   0.10 &  $236.8$ &  $1.4\pm0.5$ &   $0.82\pm0.24$ &  $1.05\pm0.30$ &  $0.16\pm0.04$ \\
23$^{*}$        &  05:38:43.5 &  -69:05:56.3 &   0.28 &  $235.8$ &  $2.7\pm0.5$ &   $0.46\pm0.11$ &  $0.85\pm0.24$ &  $0.65\pm0.07$ \\
\hline
\end{tabular}
\tablefoot{
\tablefoottext{*}{Minor axis is unresolved.}
\tablefoottext{**}{Both axes are unresolved.}
}
\end{table*}

\begin{table*}[!h]
\caption{Molecular clumps detected in the HCO$^+$ line cube.\label{tab:hco43cprops}}
\centering
\begin{tabular}{lllllllll}
\hline\hline
ID &  $\alpha$  &    $\delta$ &  $T_{peak}$ &     $v$ &   $\Delta v$ &    $R_{dc}$ &    $R$ & $L_{HCO^{+}}$ \\
 & (J2000) & (J2000) & (K) & (km s$^{-1}$) & (km s$^{-1}$) & (pc) & (pc) & (K km s$^{-1}$ pc$^2$)  \\
\hline
2         &  05:38:36.3 &  -69:05:24.2 &   3.68 &  $249.9$ &  $3.5\pm0.3$ &   $0.44\pm0.06$ &  $0.98\pm0.14$ &  $25.62\pm2.99$ \\
8$^{**}$         &  05:38:37.7 &  -69:05:33.5 &   0.13 &  $227.2$ &  $2.5\pm0.5$ &  $<0.81\pm0.21$ &  $0.69\pm0.25$ &   $0.29\pm0.04$ \\
9$^{\dagger}$        &  05:38:38.1 &  -69:05:45.3 &   0.09 &  $ 228.3 $ & - &  - &  - &  - \\
10$^{*}$        &  05:38:38.3 &  -69:05:39.6 &   0.41 &  $247.0$ &  $3.1\pm0.6$ &   $0.34\pm0.07$ &  $0.85\pm0.21$ &   $1.35\pm0.17$ \\
11$^{*}$        &  05:38:38.2 &  -69:05:41.6 &   0.09 &  $231.3$ &  $4.1\pm0.6$ &   $0.53\pm0.13$ &  $0.85\pm0.25$ &   $0.46\pm0.08$ \\
12$^{*}$        &  05:38:38.8 &  -69:05:37.3 &   0.15 &  $237.1$ &  $3.5\pm0.4$ &   $0.61\pm0.12$ &  $0.98\pm0.20$ &   $0.95\pm0.08$ \\
14$^{*}$        &  05:38:38.9 &  -69:05:46.1 &   0.12 &  $241.4$ &  $3.1\pm0.8$ &   $0.40\pm0.13$ &  $0.82\pm0.32$ &   $0.29\pm0.10$ \\
15$^{*}$        &  05:38:38.8 &  -69:05:51.1 &   0.53 &  $247.4$ &  $4.5\pm0.5$ &   $0.99\pm0.20$ &  $1.23\pm0.20$ &   $3.31\pm0.28$ \\
16        &  05:38:39.2 &  -69:05:32.3 &   2.83 &  $243.9$ &  $3.0\pm0.3$ &   $0.54\pm0.09$ &  $0.99\pm0.16$ &  $12.98\pm1.91$ \\
17       &  05:38:39.8 &  -69:05:37.6 &   1.11 &  $244.2$ &  $5.8\pm0.4$ &   $0.71\pm0.14$ &  $1.14\pm0.20$ &   $8.12\pm0.90$ \\
21$^{**}$        &  05:38:41.6 &  -69:05:25.3 &   0.11 &  $254.2$ &  $2.0\pm1.2$ &  $<0.81\pm0.30$ &  $0.56\pm0.37$ &   $0.11\pm0.08$ \\
22$^{*}$        &  05:38:41.9 &  -69:05:42.2 &   0.53 &  $237.5$ &  $2.5\pm0.3$ &   $0.82\pm0.14$ &  $1.14\pm0.17$ &   $2.86\pm0.28$ \\
23        &  05:38:43.5 &  -69:05:57.4 &   0.74 &  $235.8$ &  $2.8\pm0.3$ &   $0.52\pm0.09$ &  $0.99\pm0.17$ &   $4.23\pm0.43$ \\
\hline
\end{tabular}

\tablefoot{
\tablefoottext{*}{Minor axis is unresolved.}
\tablefoottext{**}{Both axes are unresolved.}
}
\end{table*}

\begin{table*}[!h]
\caption{Molecular clumps detected in the HCN line cube.}
\centering
\begin{tabular}{lllllllll}
\hline\hline
ID &  $\alpha$  &    $\delta$ &  $T_{peak}$ &     $v$ &   $\Delta v$ &    $R_{dc}$ &    $R$ & $L_{HCN}$ \\
 & (J2000) & (J2000) & (K) & (km s$^{-1}$) & (km s$^{-1}$) & (pc) & (pc) & (K km s$^{-1}$ pc$^2$)  \\
\hline
2$^{*}$         &  05:38:36.3 &  -69:05:24.2 &   0.87 &  $249.7$ &  $3.5\pm0.3$ &   $0.47\pm0.07$ &  $0.93\pm0.16$ &  $5.45\pm0.58$ \\
10$^{**}$        &  05:38:38.3 &  -69:05:40.0 &   0.13 &  $246.7$ &  $3.0\pm0.8$ &  $<0.81\pm0.20$ &  $0.62\pm0.25$ &  $0.29\pm0.04$ \\
15$^{\dagger}$     &  05:38:39.1 &  -69:05:52.8 &   0.09 &  $247.8$ &  - & - &  -  &  -  \\
16$^{*}$        &  05:38:39.2 &  -69:05:32.5 &   0.55 &  $244.0$ &  $2.5\pm0.3$ &   $0.36\pm0.06$ &  $0.86\pm0.18$ &  $1.98\pm0.28$ \\
17$^{*}$        &  05:38:39.7 &  -69:05:38.1 &   0.25 &  $244.2$ &  $5.0\pm0.5$ &   $0.24\pm0.04$ &  $0.77\pm0.17$ &  $1.57\pm0.14$ \\
22$^{*}$        &  05:38:42.2 &  -69:05:42.1 &   0.06 &  $237.8$ &  $2.5\pm1.0$ &   $0.65\pm0.20$ &  $0.82\pm0.31$ &  $0.13\pm0.05$ \\
23$^{*}$        &  05:38:43.4 &  -69:05:57.1 &   0.15 &  $235.5$ &  $3.1\pm0.9$ &   $0.38\pm0.09$ &  $0.83\pm0.22$ &  $0.54\pm0.07$ \\
\hline
\end{tabular}

\label{tab:hcn43cprops}

\tablefoot{
\tablefoottext{*}{Minor axis is unresolved.}
\tablefoottext{**}{Both axes are unresolved.}
}
\end{table*}

\FloatBarrier

\section{Molecular line spectra for each CO clump\label{appendix:molspectra}}

\begin{figure*}[h]
    \centering
    \includegraphics[width=0.95\textwidth]{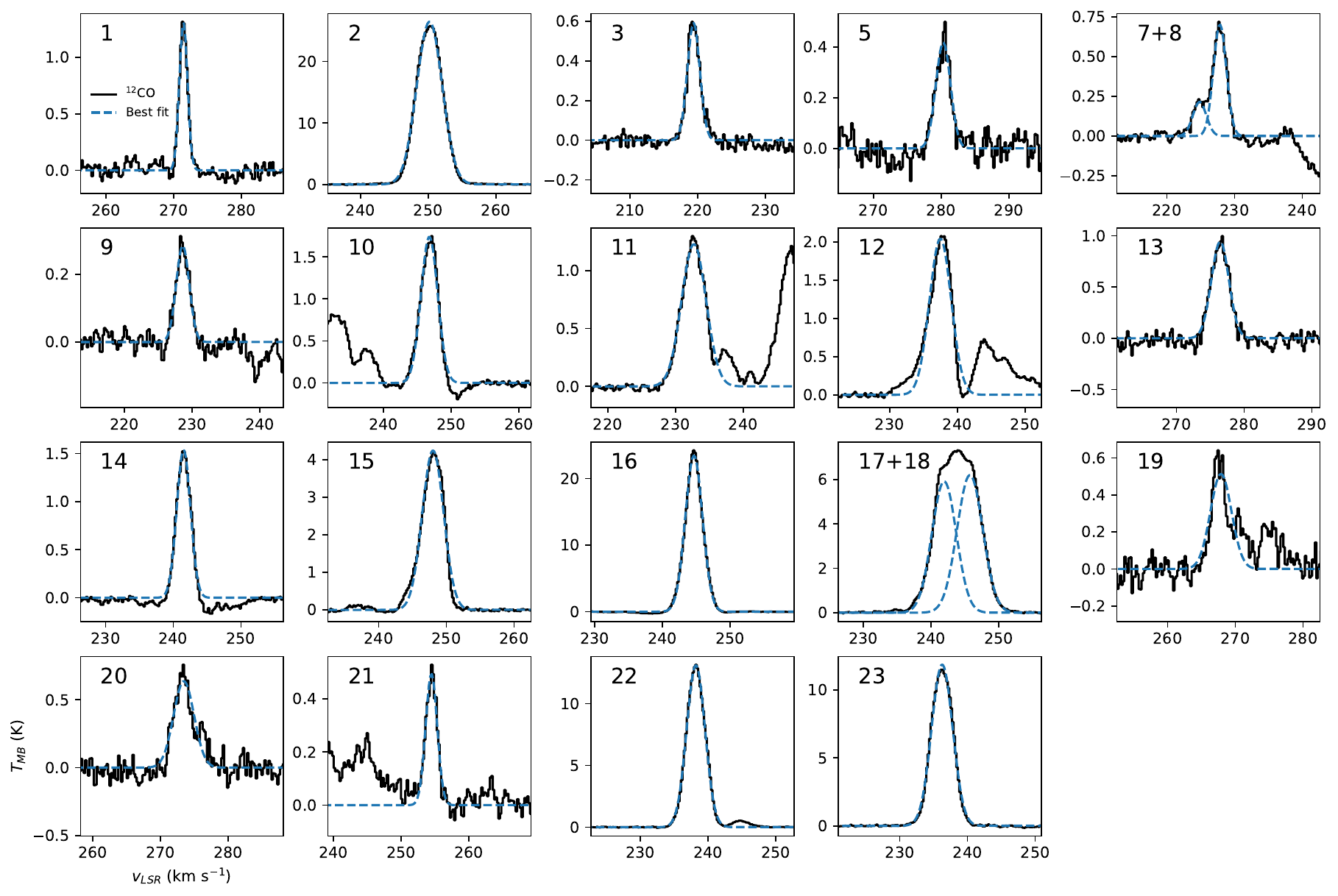}
    \caption{$^{12}$CO spectra at the peak positions of each clump. Original spectra are shown in solid black lines and the best Gaussian (or Gaussians) fit is shown in dashed colored lines. \label{fig:ap-bestfitGaussianCO}}
\end{figure*}

\begin{figure*}[h]
    \centering
    \includegraphics[width=0.95\textwidth]{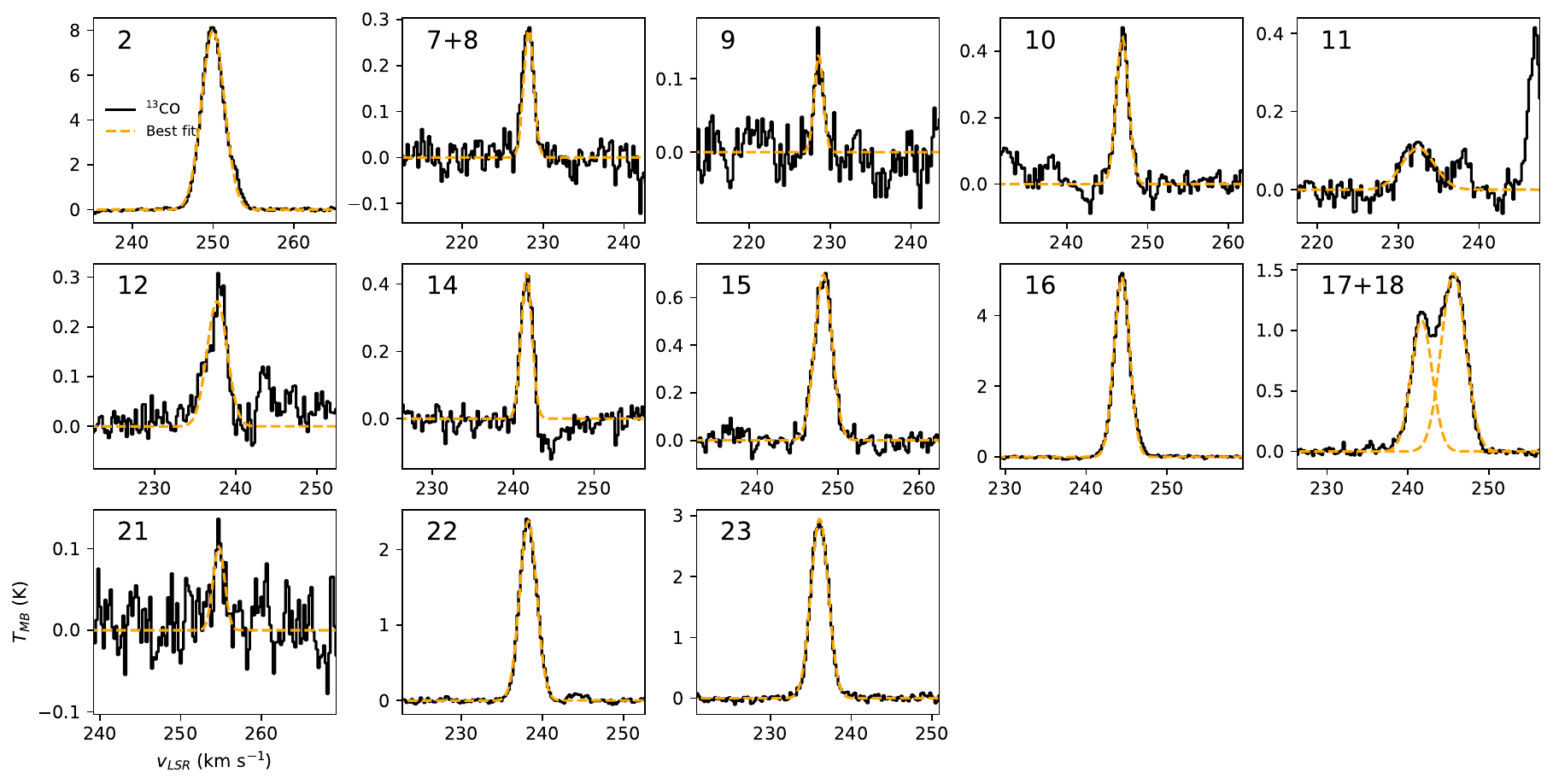}
    \caption{Same as Fig. \ref{fig:ap-bestfitGaussianCO} but for $^{13}$CO. \label{fig:ap-bestfitGaussian13CO}}
\end{figure*}

\begin{figure*}[h]
    \centering
    \includegraphics[width=0.95\textwidth]{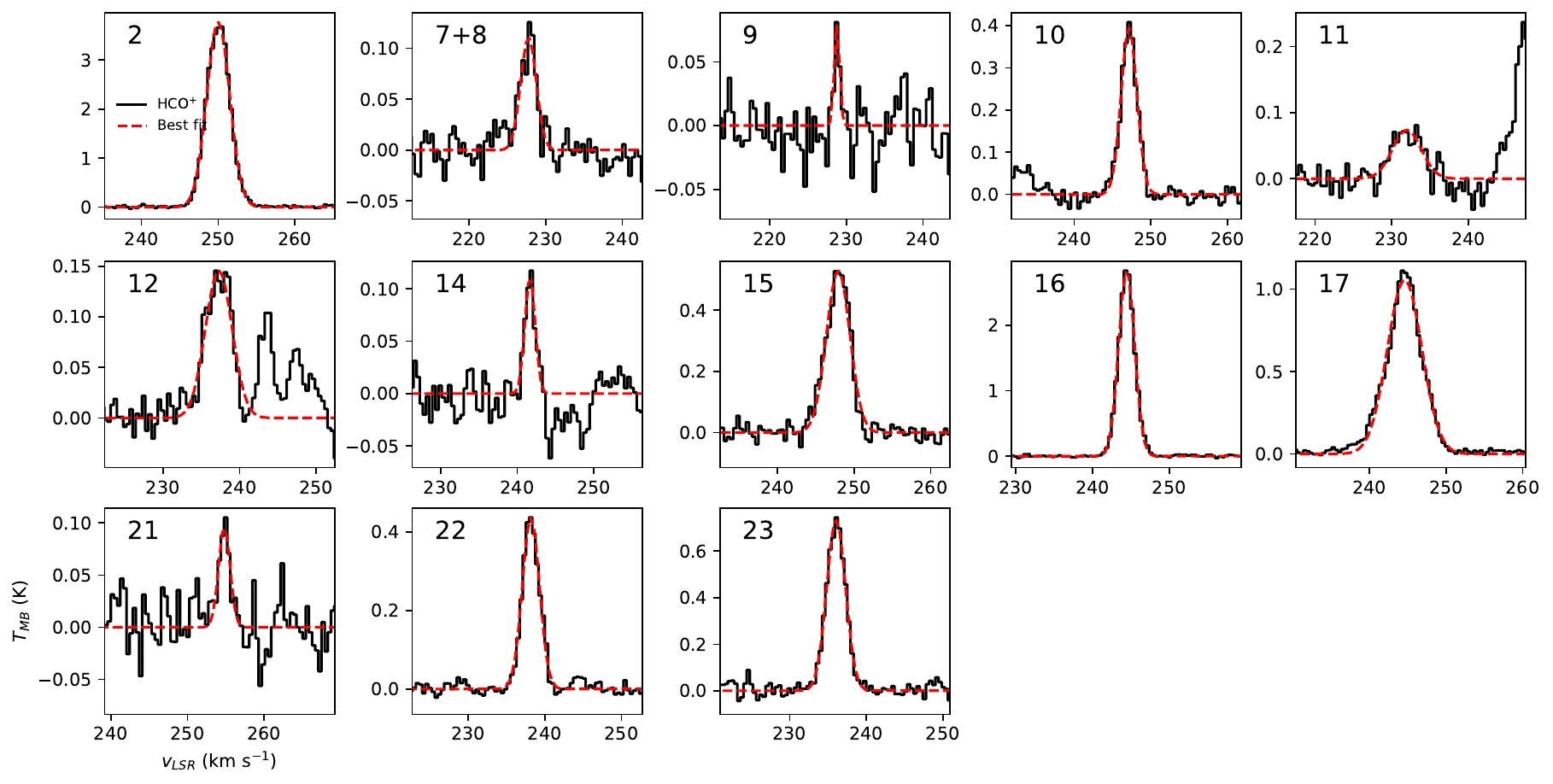}
    \caption{Same as Fig. \ref{fig:ap-bestfitGaussianCO} but for HCO$^{+}$. \label{fig:ap-bestfitGaussianHCO}}
\end{figure*}

\begin{figure*}[h]
    \centering
    \includegraphics[width=0.45\textwidth]{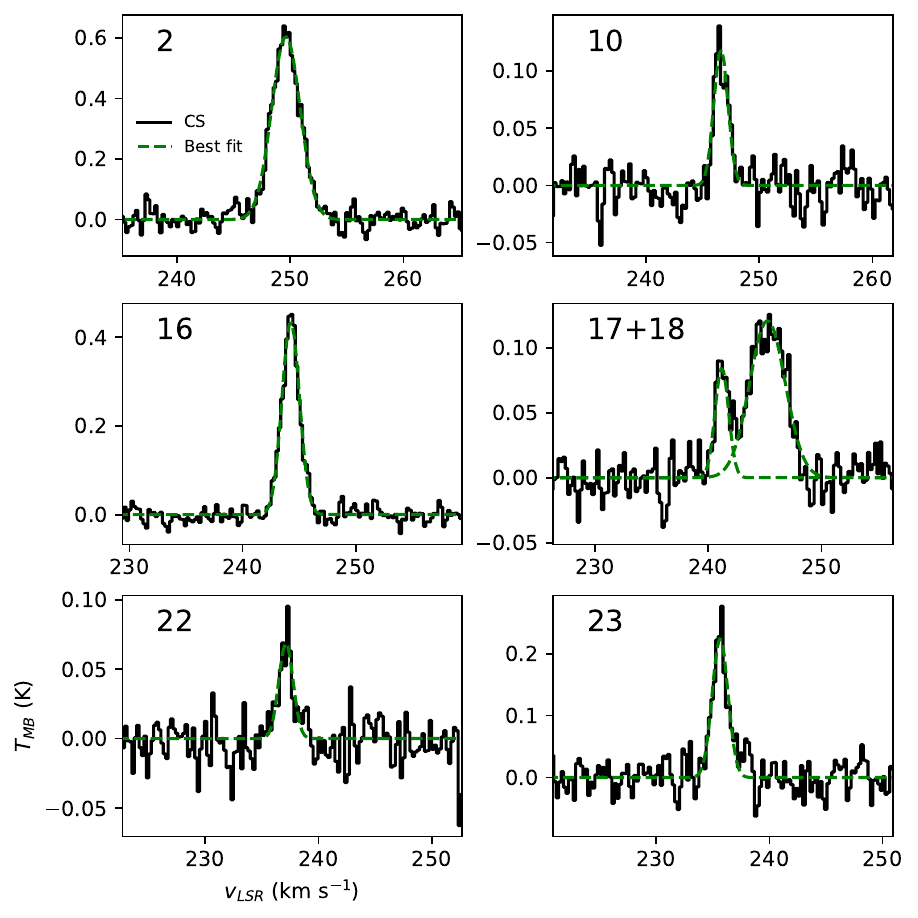}
    \includegraphics[width=0.45\textwidth]{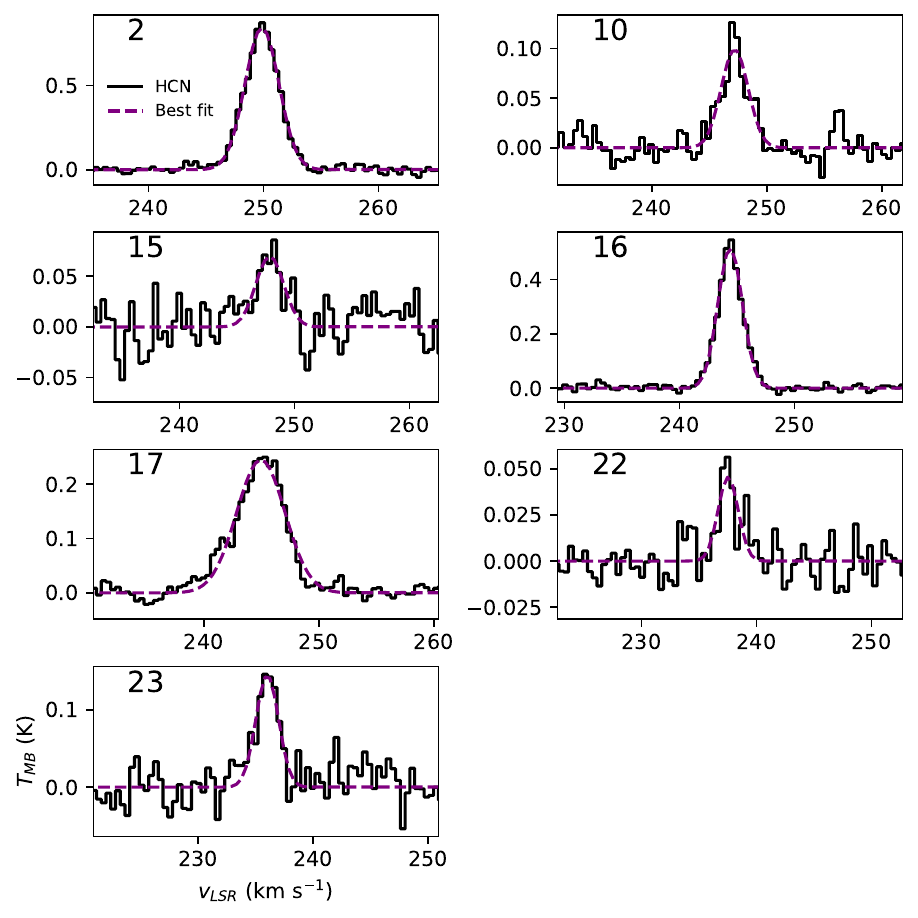}
    \caption{Same as Fig. \ref{fig:ap-bestfitGaussianCO} but for CS (left) and HCN (right). \label{fig:ap-bestfitGaussianCS}\label{fig:ap-bestfitGaussianHCN}}
\end{figure*}

Figures \ref{fig:ap-bestfitGaussianCO},  \ref{fig:ap-bestfitGaussian13CO}, \ref{fig:ap-bestfitGaussianHCO} and \ref{fig:ap-bestfitGaussianCS} show the best fit Gaussian profiles for the detected molecules in each clump, used to obtain the integrated line intensities in Sect. \ref{sec:mollinesint}. 

\end{appendix}

\end{document}